\documentclass[11pt]{article}

\usepackage[authoryear]{natbib}
\usepackage[utf8]{inputenc} 
\usepackage{abstract}
\usepackage[font={small}]{caption}
\usepackage{amsmath}
\usepackage{amsthm}
\usepackage{amssymb}
\usepackage{graphicx}
\usepackage[toc,page,titletoc]{appendix}
\usepackage{empheq}
\usepackage{subcaption}
\usepackage{float}
\usepackage{enumitem}
\usepackage[ colorlinks,
    citecolor=black,
    filecolor=black,
    linkcolor=black, urlcolor=blue]{hyperref} 
\usepackage{eurosym}
\usepackage{stackengine,graphicx}
\usepackage{parskip} 
\usepackage[bottom]{footmisc}
\usepackage{microtype}

\newcommand{\fig}{Fig.\ref}
\newcommand{\D}{\mathrm{d}}
\newcommand{\eq}{Eq.\eqref}
\newcommand{\e}{\mathrm{e}}
\newcommand{\m}{\mathrm}
\newcommand{\rs}{section \ref}
\makeatletter
\newcommand*{\centerfloat}{%
	\parindent \z@
	\leftskip \z@ \@plus 1fil \@minus \textwidth
	\rightskip\leftskip
	\parfillskip \z@skip}
\makeatother

\newtheorem{mydef}{Definition}
\newtheorem{result}{Result}
\newtheorem{mytheorem}{Theorem}
\makeatletter
\g@addto@macro\bfseries{\boldmath}
\makeatother

\title{ \bf Macroeconomic Dynamics \\ of Assets, 
Leverage and Trust } 

\author
{Jeroen Rozendaal$^1$, Yannick Malevergne$^{1,2}$, Didier Sornette$^1$}

\date{}

\begin{document}
\maketitle
\vspace{-0.7cm}
\begin{center} {\footnotesize
\emph{$^1$ ETH Zurich, Department of Management, Technology and Economics,} \\ \emph{Scheuchzerstrasse 7, CH-8092 Zurich, Switzerland} \\
$^2$ \emph{ Universit\'e de Lyon, Coactis E.A. 4161, France}\\
(Version \today)}\\
\end{center}
\hfill

\begin{abstract}
A macroeconomic model based on the economic variables (i) assets, (ii) leverage (defined as debt over asset) and (iii) trust (defined
as the maximum sustainable leverage)
is proposed to investigate the role of credit in the dynamics of economic growth, and how credit may be associated with both 
economic performance and confidence. Our first notable finding is the mechanism of reward/penalty associated with patience, as
quantified by the return on assets.  In regular economies where  the EBITA/Assets ratio is 
larger than the cost of debt, starting with a trust higher than leverage results in the highest long-term return on assets (which can be seen as a proxy for economic growth). Therefore, patient economies that first build trust and then increase leverage are positively rewarded.  
Our second main finding concerns a recommendation for the reaction of a central bank to an external shock that 
affects negatively the economic growth. We find that late policy intervention in the model economy results in the highest long-term return on assets and largest asset value. But this comes at the cost of suffering longer from the crisis until the intervention occurs. 
The phenomenon that late intervention is most effective to attain a high long-term return 
on assets can be ascribed to the fact that postponing intervention allows trust to increase first, and it is 
most effective to intervene when trust is high.
These results derive from two fundamental assumptions underlying our model: (a) trust tends to increase when it is above leverage;
(b) economic agents learn optimally to adjust debt 
for a given level of trust and amount of assets. Using a Markov Switching Model for the EBITA/Assets ratio,
we have successfully calibrated our model to the empirical data of the return on equity of the EURO STOXX 50 for the time period 2000-2013.
We find that dynamics of leverage and trust can be highly non-monotonous with
curved trajectories, as a result of the nonlinear coupling between the variables. This has an important implication for policy makers, 
suggesting that simple linear forecasting can be deceiving in some regimes and may lead to
inappropriate policy decisions. 

\end{abstract}


\section{Introduction}

The credit crisis and panic that erupted in 2008 in the US and spilled over 
the World have made again clear that financial price and economic value
are based fundamentally on trust; not on fancy mathematical formulas, not on subtle
self-consistent efficient economic equilibrium; but on trust in the future, trust in
economic growth, trust in the ability of debtors to face their liabilities, trust in
financial institutions to play their role as multipliers of economic growth, trust that
our money in a bank account can be redeemed at any time we choose. Usually, we
take these facts for granted. When depositors happen to doubt banks, this 
leads to devastating bank runs. When banks start to doubt other banks, this leads
to a freeze of the inter-banking loan markets and an effective run on collaterised assets.
Then, the implicit processes of a working economy --all we take for granted-- starts to dysfunction and spirals into a
global collapse, as almost happened with the Lehmann Brothers Bankruptcy.
The standard discourse by observers and pundits is to attribute the 
2008 crisis to the mortgage-backed
securities linked to the bursting of the house price bubble, the irresponsible lending,
overly complex financial instruments and conflicts of interest leading to asymmetric
information translated into market illiquidity, and the spreading of risks via packaging
and selling of imagined valuations to unsuspecting investors. What is missing in this 
discussion is to endogenise trust in the dynamics of the credit system.

As a contribution to fill this gap, 
we construct a simple macroeconomic model, based on basic accounting rules combined with reasonable economic assumptions,
which put trust as the central dynamical variable. Focusing on credit creation and its dynamics, 
we naturally use asset amount and leverage (here defined as the ratio debt to asset) as the two other variables.
To operationalise the model, we define trust as the maximum sustainable leverage. This transforms
an a priori ill-defined qualitative concept, capturing the degree of belief in the reliability, truth, or ability of something,
into a quantitative variable that can be worked with.

Our motivation to focus on  the three variables (i) assets, (ii) leverage and (iii) trust is that these are three key economic variables for studying different economic regimes, since they  relate to credit, which has a central role in economics as a driver of economic cycles. For example, credit is pivotal to understand the financial crisis of 2008. Ahead of  the  2008 crisis, credit was easily available: a high percentage of asset value could be used as collateral to obtain a loan. The high availability of credit caused debt levels and leverage to increase rapidly.  At a certain moment, this drastically changed and credit crunched: the financial crisis of 2008 was a fact. To exemplify the importance of credit today: the current global public debt is over 57 trillion U.S. dollar \citep{Economist}. For comparison, the Gross World Product (GWP) is about 78 trillion U.S. dollar \citep{WorldBank}.
For  the United States, the public debt as a percentage of GDP is over 90\% at the time of writing.
The central role of credit (debt) in an economy suggests to focus on the modelling the
interplay between assets (assets serve as a collateral to obtain a loan), leverage (the debt to assets value) and trust (the maximum sustainable leverage). 

The link between credit and the afore-mentioned variables has been qualitatively discussed by \citet{VonderBecke}, who
provided  a qualitative framework in which credit creation is argued to depend on the amount of collateral assets accepted, the level of leverage and the level of trust and confidence in future cash flows. In their paper, a qualitative theory was proposed to understand  credit creation and the perspectives of different schools of thought were integrated (Austrian, Mainstream and Post Keynesian).  
Besides the paper of \citet{VonderBecke}, there are numerous other papers in which assets, leverage and trust are (mostly separately) studied.

\citet{Geanakoplos} presents a model in which the interplay of leverage and assets is captured. In this model, a set-up is used in which houses are used as collateral for  long-term  loans, and loans are used as collateral for repos\footnote{Abbreviation for repurchase agreements, i.e. short-term collateralized loans.}. 
 Based on this framework, \citet{Geanakoplos} advocates that leverage rates should be the primary target (instead of interest rates) of central banks in times of crisis.

 \emph{Asset} prices are  a central theme in, for example, the research of \citet{Bernanke}. Their paper studies  whether central banks should respond to movements in asset prices.
\citet{Borio} linked price changes of assets to financial instability: sustained rapid credit growth combined with large increases in asset prices tends to increase the probability of financial instability. Based on historical data, the research showed that significant changes in asset prices are linked to increased financial instability. It is argued that monetary policy should respond to changing asset prices with the goal of preserving financial stability.

\citet{Lang} present a study in which \emph{leverage} is the main variable that is studied and it is demonstrated that an inverse  relation between leverage and future growth of firms with a low q-ratio\footnote{Tobin's q-ratio is defined as the  total market value of the firm divided by the the total asset value.} exists. 
Leverage as an indicator for growth opportunities is also a theme in a study of \citet{Gilchrist}, who hypothesizes that high leverage economies are particularly vulnerable to slowdowns in the world economy.

\citet{Putnam} initiated the study of \emph{trust} (``social capital'') and its relation to the well-functioning of a society by providing case studies of the functioning of regional governments in Italy.
\citet{Knack} have investigated trust and its impact on the economy. Survey data was used to capture trust and it was found that ``social capital'' (indicated by trust) matters  for economic performance. 
 \citet{Dincer} confirm this finding based on data from U.S. states in which a positive relation between trust and economic growth was found.
In a study of \citet{Bjornskov}, it has been investigated through what mechanisms trust affects economic growth and trust is proposed to be the fundamental driver of economic development. 
There is no full consensus on the relation between trust and economic growth. \citet{Beugelsdijk} could not verify the link between trust and economic growth and they concluded that there may not always be an economic pay-off of trust. Hence, there exists no consensus on a relation between trust and economic performance and one should keep in mind that the definition of trust may differ in various papers.

In order to position our model, it is useful to provide a short overview of
influential macroeconomic models in the academic and central banking literature. This overview is useful to to position our assets, leverage
and trust model.

The New-Keynesian dynamic stochastic general equilibrium  (DSGE) model is one of the most influential models in the academic and central banking community (see e.g. \citet{Sbordone} for an overview of DSGE models and see  \citet{Isohatala} for an overview of DSGE models and beyond). The model is built around three coupled equations, which derive from micro-foundations: supply, demand and monetary policy equations. Furthermore, equations describing expectations can be explicitly taken into account in the DSGE framework. A strong element of the model is that it can capture the response of output and inflation to demand, supply and policy shocks. DSGE models however failed to explain the significant rapid changes (declines) in  asset prices, output and investment that occurred during the 2008 crisis and thereafter \citep{Isohatala,Reinhart}.

As a response to this inability to model the 2008 crisis, a strand of the recent literature focussed on introducing dynamics to account for the possible occurrence of extreme events that abruptly and significantly influence economic quantities such as asset prices, output or investment.  Examples of influential recent papers that focus on this are those by \citet{He} and \citet{Brunnermeier}. The new models employ continuous time modelling approaches to macroeconomic problems  and assume market incompleteness (i.e. not all risks can be hedged). This  approach is mathematically convenient as it allows one to characterise the possible economic states by (partial) differential equations.
A characteristic of the new class of models is that, in the limit of small (zero) disturbances/volatility, the non-linear dynamics of the model reduces to a linearised DSGE model.

Variants of the afore-mentioned models,  as well as the model constructed in this research, can eventually be used to assess policy measures of central banks.
 \citet{Bacchetta} focus on the question whether monetary policy can help to avert self-fulfilling debt crises and at what cost (particularly in terms of inflation). 
 Their paper builds on earlier sovereign debt crisis models and extends  these with the goal of quantifying the afore-mentioned cost.
 A model on which it builds is that of \citet{Lorenzoni} who introduced a real sovereign debt crisis model (without a monetary authority), in which a government defaults at a predetermined time $T$ if the present value of debt exceeds the present value  of future primary surpluses (i.e.  government spending, excluding  interest payments, minus  income from taxes). \citet{Bacchetta}  extend the model to a monetary economy, which provides a convenient framework to study  conventional monetary policies (in particular the effects of inflation and interest rate). Furthermore, it allows to study additional (non conventional) monetary tools, such as buying government bonds by considering the budget constraint of a central bank. \citet{Bacchetta} assess the conventional monetary policies (operating through interest rates) using a New Keynesian DSGE model based on \citet{Gali} with extensions by \citet{Woodford} to introduce a delay in the impact of monetary policy shocks. 
  \citet{Bacchetta} conclude that the conventional tools of the central bank, aimed at averting a debt crisis, lead to high inflation for a sustained period of time. Unconventional monetary policies (e.g. quantitative easing: the purchase of government securities or other securities from the market) are suggested to be effective only when an economy is at the zero lower bound (ZLB), i.e. when the short-term interest rate is  at 0.

 The organisation of the present article is as following. Section \ref{sec:formalisation_dynamics} gives the formal definition of our three
 fundamental variables as well as a number of derived quantities. It then states the three governing ordinary differential
 equations, whose derivation is presented in Appendix \ref{app:derivation_dynamcis} and some closed-formed solutions are given. The fixed point
 stability analysis is also presented. Section \ref{sec:results_fulldynamics} presents a survey of the main properties of the model, by 
 showing phase portraits of the trajectories in the leverage-trust space and by quantifying the associated return on asset.
 Section \ref{sec:leverage_trust_regime_shifts} uses the model to investigate what happens under regime shifts, either due to some exogenous adverse shock leading to a sustained crisis (negative growth rate) and/or as a result of policy intervention in the form of decreased 
 target interest rates and increased return on assets. Section \ref{sec:Bayesian} presents the calibration of the model, extended
 using a Markov Switching framework for an exogenous model parameter, to the return on equity of the EURO STOXX 50 for the time period 2000-2013.
 Section \ref{sec:conclusion} concludes. The derivations and proofs of the main results are presented in Appendices \ref{leverage_trust_appendix} and \ref{app_proof_theorem}.

\section{Formalisation of the joint dynamics of assets, leverage and trust} \label{sec:formalisation_dynamics}

\subsection{Basic definitions: assets, leverage, trust}

The following definitions clarify the meaning of
our three key economic variables and introduce their growth rates.\\

\begin{mydef}
The total asset value at time $t$ is denoted by $A(t)$. Asset value can always be written as the sum of debt and equity:
\begin{flalign}
A(t) := D(t)+ E(t), \label{assets_sum_debt_equity}
\end{flalign}
where $D(t)$ is the debt at time $t$, and $E(t)$ is the equity at time $t$.
\label{def_assets_sum}
\end{mydef}

\hfill

\begin{mydef}
The leverage at time $t$, denoted $L(t)$, is defined here as:
\begin{equation}
L(t):= D(t)/A(t). \label{Leverage_def}
\end{equation}
\end{mydef}
 Banks usually operate with leverage ratios close to 1, while ``normal firms'' usually operate with  lower leverage ratios. The leverage is an interesting quantity to examine since it quantifies the indebtedness of an economy and is  linked to risk and volatility. As a side note, the standard way of defining leverage is $L_s:=D/E$. For this study, the definition given by \eq{Leverage_def} is more convenient to link to the trust variable
 to be defined formally in definition \ref{constraint_for_debt}.
 $L$  in \eq{Leverage_def} is related to $L_s$ through $L = \frac{L_s}{1+L_s}$.
While $L_s$ has no limit in principle, $L$ has an upper limit of one, corresponding to the limit of zero equity ($E=0$).

\hfill

\begin{mydef}
The trust $T(t)$ is defined as the fraction of the total assets that qualifies as collateral for taking on debt. The trust can hence be viewed to represent a borrower's creditworthiness. 
It is assumed that under normal circumstances the level of trust $T(t)$ satisfies the following inequality:
\begin{equation}
D(t) \le T(t) \cdot A(t).  \label{constraint_for_debt}
\end{equation}
This inequality expresses that in general the debt $D$ should not exceed the total value of acceptable collateral.
This holds simply by definition of what is meant by ``acceptable collateral'' and derives from the nature of lending
where the lender hedges his risks by ensuring that the borrower has assets at least as large as the borrowed amount.
\end{mydef}
We should note that periods of market exuberance have been characterised by a failure of this 
inequality, as for instance for subprime  ``NINJA loans'' (loans extended to people with ``No Income, No Job, (and) no Assets).

\hfill

\begin{mydef}
 In this study, the return on assets (ROA), defined by
 \begin{equation}
r_{A} (t):= \frac{1}{A} \frac{\D A}{\D t}~, \label{return_assets_def}
\end{equation}
 is viewed as a proxy for  economic growth.
 \end{mydef}
 We will also use the growth rate $r_D  (t) := \frac{1}{D} \frac{\D D}{\D t}$ of debt,
 the return $r_{E} (t):= \frac{1}{E} \frac{\D E}{\D t}$ on equity (ROE) 
 and the growth rate $r_L (t) :=\frac{1}{L} \frac{\D L}{\D t}$ of leverage.

From differentiation of $\ln(D) = \ln(L) + \ln(A)$ (\eq{Leverage_def}) with respect to $t$,
the growth rate $r_L (t)$ of leverage, the growth rate $r_D (t)$ of debt  and the return $r_A (t)$ on assets are linked by
the simple relation
\begin{equation}
r_D (t) =r_L (t) + r_{A} (t).  \label{debt_lev_assets_growth}
\end{equation}

\subsection{System equations for the joint dynamics of assets, leverage and trust \label{app:derivation_dynamcis}}

We now derive the coupled dynamics of assets, leverage and trust.

\subsubsection{Auxiliary definitions} \label{sec:notation}
Before presenting the fundamental equations for assets, debt and trust, it is necessary to introduce additional  definitions. 

\begin{mydef}
Depreciations $\mathcal{D}_{\D t} (t)$ (over a period $\D t$) are given by:
\begin{flalign}
\mathcal{D}_{\D t} (t) =  \delta_{\D t} A(t),
\end{flalign}
where $ \delta_{\D t} $ is a depreciation factor, indicating what fraction of asset value is ``lost'' over a time period of $\D t$.
\end{mydef}

\hfill

\begin{mydef}
Amortization  $\mathcal{A}_{\D t} (t)$ (over a period $\D t$) is given by:
\begin{flalign}
\mathcal{A}_{\mathrm{\D t}} (t)=  D(t-\D t)-D(t),
\end{flalign}
where $\mathcal{A}_{\mathrm{\D t}} (t)>0$ if the debt decreases (the word derives from ``amortisen'', which means ``to kill'': decrease debt in this case). Note that $\mathcal{A}_{\mathrm{\D t}}(t)<0$ if  more debt is taken on.
\end{mydef}

\hfill

\begin{mydef}
    The definition of net income (NI), also referred to as net earnings, $\mathcal{E}_{\D t} (t) $ follows from  accounting:
\begin{flalign}
\mathcal{E}_{\D t} (t) = \underbrace{\underbrace{ \kappa_{\D t} A(t)}_{\mathrm{EBITDA}}- \underbrace{ \delta_{\D t}  A(t)}_{\mathrm{Depreciations}} -\underbrace{[ D(t-\D t)-D(t)]}_{\mathrm{Amortization}}}_{\mathrm{EBIT}}- \underbrace{ r_{\D t} D(t)}_{\mathrm{Interests \; payments} }  , \label{earnings}
\end{flalign}
in which  $ \kappa_{\D t}$  is  defined as the EBITDA to assets ratio. Furthermore, $ r_{\D t}$ is the interest rate paid on debt. 
The subscripts $\D t$ refers to a period $\D t$ (so e.g. $\mathcal{E}_{\D t} (t)$ are the net earnings generated over a time period $\D t$). Note that taxes are neglected in the model (in case of taxes, these should be subtracted from EBIT as well in \eq{earnings}; such that NI=EBIT-Interest-Taxes)
\end{mydef}

\noindent Furthermore, the net earnings can be classified by how they are allocated: they are  either paid out as dividends to the shareholders, or reinvested into the company and/or maintained as cash.
\begin{flalign}
&\mathcal{E}_{\D t} (t) = \underbrace{ p^{\mathrm{out}}  \mathcal{E}_{\D t} (t)}_{ \mathrm{Dividends}} + \underbrace{ p^{\mathrm{back}}   \mathcal{E}_{\D t} (t)}_{\mathrm{Retained \; Earnings}}, 
\end{flalign}
where $p^{\mathrm{out}} $ is the pay out ratio and $p^{\mathrm{back}} $ is the plow back ratio. Note that: $p^{\mathrm{out}}+p^{\mathrm{back}} =1$. In the next sections, it is assumed that $p^{\mathrm{back}}=1$, and hence the net earnings are equal to the retained earnings. This assumption can be motivated by invoking the Modigliani–Miller theorem \citep{MM}: a firm's dividend policy is irrelevant for the valuation of shares, under idealized conditions which include the absence of taxation, transaction costs, asymmetric information and market imperfections.

\subsubsection{Assets} \label{sec:assets}

From basic accounting under the assumption of a plow back ratio of 1, it follows that:
\begin{equation}
  {\footnotesize \m{Asset \; value}(t)= \m{Asset \; value}(t-\D t)+ \m{Net \; Income}(t)}.
\end{equation} 

\noindent Using the notation as introduced in \rs{sec:notation}:
\begin{flalign}
 A(t) &= \; A(t-\D t) + \mathcal{E}_{\D t} (t), \label{accounting_eq} \\
&\stackrel{\mathmakebox[\widthof{=}]{\eqref{earnings}}}{=} 
\; A(t-\D t) +  \underbrace{( \kappa_{\D t}-     \delta_{\D t} ) A(t)}_{\m{EBITA}}    -\underbrace{[ D(t-\D t)- D(t)]}_{\m{Amortization}}-  \underbrace{r_{\D t} D(t)}_{\m{Interest \; payments}}.\label{accounting_eqII}
\end{flalign}

By defining $g_{\D t}$ to be the EBITA/Assets ratio, i.e.  $g_{\D t}:=  \kappa_{\D t} -  \delta_{\D t}$, \eq{accounting_eqII} can be rewritten as follows:
\begin{equation}
  {A(t) - A(t-\D t)} = {  g_{\D t}  A(t) } -{ r_{\D t} D(t)} + [D(t)-D(t-\D t)]. \label{accounting_eq2}
\end{equation}

\noindent By dividing \eq{accounting_eq2} by $\D t$ and taking the limit $\D t \to 0$, one obtains:
\begin{flalign}
 \frac{\D A}{\D t} = g A(t) -r D(t)+ \frac{\D D}{\D t},  \label{assets1}
\end{flalign}
where $\displaystyle g := \lim_{\D t \to 0} \frac{ g_{\D t}}{\D t} $ and $\displaystyle r:= \lim_{\D t \to 0} \frac{  r_{\D t}}{\D t}$.  
 \eq{assets1} is the the governing equation for assets.

\subsubsection{Trust} \label{sec:trust}

To construct a fundamental equation for trust, we firstly assume that in the absence of externalities (that might cause abrupt changes in the trust) the trust increases to its maximum value $T=1$, where all assets qualify as collateral.
A tendency for trust to increase is thus assumed. 
This assumption can be supported  by both a psychological and a rational phenomenon.
The psychological phenomenon is  the prosociality  of humans: people have a tendency of helping, benefiting and trusting others and/or society \citep{Dovidio}.
 The rational phenomenon is the aim to use all assets and let the economy function at its ``full potential''. In an economy that functions to its full potential, the trust is at its maximum. This is favourable because the higher the trust, the  more credit there can be  to fund profitable projects (if trust is 1, 100\% of the asset value is accepted as a collateral to obtain a loan). 

 A convenient and common function to capture the growth of the trust to its maximum 1, is the logistic function (``S-curve''). It was first introduced  by Pierre Verhulst in the nineteenth century to describe population growth, but it has a wide range of applications in many fields. For example, it has  applications  in statistical physics (Fermi-Dirac), demography, and to model the dynamics of the market penetration of a product or service. 
A standard logistic equation for trust would thus be:
\begin{equation}
\frac{\D T}{\D t} =k T(1-T), \label{trust_equation2}
\end{equation}
where $1/k>0$ denotes the inertia (resistance) of the trust to change. \eq{trust_equation2} captures the standard features of a logistic equation. For low $T$, \eq{trust_equation2}  simplifies to $\displaystyle \frac{\D T}{\D t} \approx k T$, which expresses that the growth rate of trust is proportional to the existing trust.  This type of accumulation of trust is named the Matthew effect \citep{Merton}. In this context it follows that,  the higher the level of existing trust, the faster  trust grows  (``cumulative advantage''). For $T$ close to 1 (large $T$),  \eq{trust_equation2} simplifies to  $\displaystyle \frac{\D T}{\D t} \approx k (1-T)$, which  shows that the rate of change of trust ($\frac{\D T}{\D t}$) decreases when $T$ approaches 1. Note that when $T \to 1$ or $T\to 0$ then a steady-state is reached ($\frac{\D T}{\D t} \to 0$ in \eq{trust_equation2}). 
 
It is desired that the growth of trust is positive when trust exceeds leverage, while it should be negative when leverage exceeds trust. In this way, a convergence to the ``natural optimum'' $T=L$ is imposed. 
 By adding a multiplication of $(T-L)$ to \eq{trust_equation2}, this feature is captured. The following ``extended'' logistic equation is hence postulated for the trust:
\begin{flalign}
\frac{\D T}{\D t} = k (T-L) T(1-T). \label{trust_equation}
\end{flalign}
Another way to justify the term $(T-L)$ is to argue that the growth rate $k$ of trust should be modulated by $(T-L)$, being positive if $T>L$ and negative if 
$T<L$. A first order Taylor expansion then yields the form leading to expression (\ref{trust_equation}).
 The ``extended'' logistic equation (\eq{trust_equation}) is more complex than the standard logistic equation (\eq{trust_equation2}) because a constant trust (steady-state) can be reached when either $L\to T$, $T \to 1$, or $T\to 0$ which ever occurs first. Given that we start from an initial condition $T(0)>L(0)$, this means that if $L$ ``catches up'' with $T$ before $T \to 1$, a stationary condition is reached where $T_{\text{stationary}} =L_{\text{stationary}}$.

\subsubsection{Debt} \label{sec:debt}

Based on the assumption of a tendency for economies  to increase debt levels, as argued by  e.g. \citet{Dalio} and empirically supported by \citet{Graeber}, a debt dynamics will be proposed in this subsection. We assume a tendency for economies to reach its maximal  debt, while taking into account \eq{constraint_for_debt}. This implies, in mathematical terms,  that  $D=TA$ should be imposed to be a natural fixed point. 

Then, the next point of reasoning is that a non-instantaneous  convergence of $D\to TA$ is assumed. There exists  inertia for $D$ to reach $TA$ because it takes time to find investment opportunities and it only makes sense to increase debt when such opportunities exist.  This can be supported by, for example,  \citet{Taylor} and \citet{DeAngelo}.  \citet{Taylor} discusses that there are occurrences in history where the capital mobility (moving funds internationally) was found to be low.  \citet{DeAngelo} discuss leverage ratios with slow average speeds of adjustment to target level.  
  
The  governing equations for the debt is postulated to be: 
\begin{flalign}
& \frac{\D (D-TA)}{\D t}= -a ({D - TA }), \label{Debt_dv} \\
\Leftrightarrow \; & \frac{\D D}{\D t} = a (TA- D)+ \frac{\D (TA)}{\D t}, \label{Debt1}
\end{flalign}
where $1/a>0$  captures the inertia of debt convergence to its optimal value. Note that \eq{Debt1} is valid for $D<TA$ and $D>TA$: in both cases there is a tendency for $D \to TA$. 

\noindent As previously mentioned, \eq{Debt1} is constructed such that $D$  tends to converge to $TA$.  This convergence of $D$ to $TA$ would already be ensured without the term $\frac{\D (TA)}{\D t}$ in \eq{Debt1}, however this term can boost or slow down the change in $D$, depending on the rate of change (trend) in $TA$. The proposed debt equation (\eq{Debt1}) assumes that this trend of $TA$ is known and incorporates this. We assume that economic agents learn optimally to adjust debt for a  given level of trust and amount of assets.

\subsubsection{Final equations for assets, leverage, trust} \label{appendix_final_ALT}

In this section, we outline how the final differential equations for assets, leverage, trust can be  obtained. 

First of all, \eq{assets1}, \eqref{trust_equation},  \eqref{Debt1} are divided by $k$, thereby making the equations non-dimensional.  We introduce the non-dimensional time $\tau$, which we define as follows:
\begin{equation}
\tau := kt.
\end{equation}
We denote non-dimensional model parameters ($a, g, r$) with a tilde, for example: $\tilde a: = a/k$.

 By substitution of  \eq{Debt1}, and then \eq{trust_equation}, in   \eq{assets1} we  obtain the  final assets equation:
\begin{equation}
 \frac{\D A}{\D \tau } = \left(\frac{\tilde g- \tilde r L+ \tilde a (T-L) }{1-T} + (T-L) T \right) A, \label{A_tau_appendix}
\end{equation}

The final leverage equation can be obtained by using the definition of leverage (\eq{Leverage_def}) and the trust equation (\eq{trust_equation}):
\begin{equation}
\frac{\D L}{\D \tau}= (T- L)    \left(\frac{\tilde g-\tilde r L+\tilde a (1-L) }{1-T} + (1-L) T \right), \label{L_tau_appendix}
\end{equation}
which can also be written as follows:
\begin{flalign}
\frac{\D L}{\D \tau}= (T-L) \left(\beta\frac{L_0 - L  }{1-T}+ T(1-L)\right), 
\label{L_tau_2} 
\end{flalign}
where $\beta$ and $L_0$\footnote{$L_0$ appears naturally when substituting \eq{Debt1} into \eq{assets1} for $T=1$, and then solving for $L$.} are defined as follows: $ \displaystyle \beta :=\frac{a+r}{k} = \tilde a + \tilde r,  \;  L_0 := \frac{g+a}{r+a}$. \eq{L_tau_2} is a convenient expression for the study of fixed points.

Lastly, the final trust equation is simply the non-dimensional version of \eq{trust_equation}:
\begin{equation}
\frac{\D T}{\D \tau} = T(T-L)(1-T). \label{T_tau_appendix}
\end{equation}

\hfill

Summing up, the assets, leverage and trust variables are governed by following system of three coupled ordinary differential equations:
\begin{empheq}[left = \empheqlbrace]{align}
& \frac{\D A}{\D \tau } = \left(\frac{\tilde g- \tilde r L+ \tilde a (T-L) }{1-T} + (T-L) T \right) A, \label{A_tau}  \\
& \frac{\D L}{\D \tau}= (T- L)    \left(\frac{\tilde g-\tilde r L+\tilde a (1-L) }{1-T} + (1-L) T \right), \label{L_tau} \\
&  \frac{\D T}{\D \tau}=   T(T-L)  (1-T), \label{T_tau}
\end{empheq}
where  $\tau$ represents a non-dimensional time expressed in unit of the characteristic
time scale $1/k$ defined in the trust dynamics (\ref{trust_equation2}). The
tildes on the model parameters ($\tilde a, \tilde g, \tilde r$) indicate that the paramters $a, g$ and $r$ 
introduced above have also been normalised by $k$. 
The meaning of $\tilde a, \tilde g$ and $\tilde r$ are the following:
(i) $1/\tilde a$ is the characteristic time scale for the debt to reach its optimal value ($D \to TA$);
(ii) $\tilde g$  is the EBITA/Assets ratio, and (iii) $\tilde r$ is the interest rate paid on debt.

From the above assets, leverage, trust equations, it can be observed that the leverage and trust equations constitute a sub-system (independent of $A$).

\subsection{Closed-form solutions of the dynamics}

The ROA can be obtained from \eq{A_tau} and the ROE can be obtained from Definition \ref{def_assets_sum} and the fundamental assets equation as presented in section \ref{sec:assets} (i.e. \eq{assets1}). 

 \hfill
 
 \begin{result}
 The ROA and ROE as a function of trust and leverage, and dependent on the model parameters $\tilde a, \tilde g, \tilde r$, are given by:
\begin{align} 
&\tilde r_{A} = \tilde g \frac{1}{1-T} -\tilde r \frac{L}{1-T} + \tilde a \frac{T-L}{1-T} + (T-L)T, \label{r_A_general2} \\
& \tilde r_E = \tilde g + \frac{L}{1-L} (\tilde g-\tilde r).  \label{r_E_expr}
\end{align}
\end{result}

From \eq{r_A_general2},  the different contributions to the ROA  can be  seen. 
The first term in the r.h.s. is the EBITA/Assets ratio leveraged by trust (i.e. $ \propto \frac{1}{1-T}$).
The second term is the cost of debt leveraged by trust. The third term, which is also leveraged by trust, is positive
(resp. negative) for $T>L$ (resp. $T<L$). It can be interpreted as a reward
(resp. penalty) from being patient (resp. impatient) by first building trust and then increasing leverage (resp. by 
increasing leverage before establishing trust). The fourth term is the transient economic growth (resp. contraction) 
resulting from the catching up of an economy that grows its leverage towards its optimal value from below (resp. above).

In \eq{r_E_expr} for the ROE, we see that the term $(\tilde g - \tilde r)$ is leveraged by how close leverage
is to its maximum value $1$.

The leverage/trust trajectories can be determined analytically based on \eq{L_tau} and \eqref{T_tau}. 
The results derived in Appendix \ref{leverage_trust_appendix}  can be summarised as follows.

\hfill

\begin{result}
The leverage/trust trajectories are given by:
\begin{align}
L(T) &= 1 - K\frac{(1-T)^{1+\beta}}{T^\beta} \e^{-\frac{\beta}{1-T}} 
 +  (L_0 -1)\bigg\{ \left[ \frac{\beta}{1+\beta}+ \frac{1-T}{1+\beta}\right] \notag \\
 & \;\;\;+ \frac{\beta}{1+\beta}  \frac{T^2 }{(1-T)}  \int_0^1{(1-y)^{\beta+1} } {\e^{-\frac{\beta T}{1-T} y } \D y } \bigg\},    \label{LT_final_th}
\end{align}
where $K$ is an integration constant, $\beta:=\tilde r + \tilde a$, and  
\begin{equation}
L_0:= \frac{\tilde g +\tilde a}{\tilde r+ \tilde a}~.
\label{wththtrgqfq}
\end{equation}
\label{Result_LT_trajectories}
\end{result}

\eq{LT_final_th} is a closed-form analytical solution. It is also useful to use for instance an Euler discretisation scheme
to study numerically the different regimes described by  \eq{L_tau} and \eqref{T_tau}.

\subsection{Analysis of the leverage and trust subsystem (fixed points and stability)}

The following theorems  \ref{FixedPoints_determ}-\ref{Theorem_fixedpoints2} present 
important properties of the system of equations as presented in \rs{appendix_final_ALT}. Their proof 
is provided in Appendix \ref{app_proof_theorem}.

\hfill

\begin{mytheorem}
The fixed points of the leverage/trust subsystem are the points $(T, L)=(0,L_0)$ and $(T, L)=(1,L_0)$, 
where $L_0$ is defined by expression (\ref{wththtrgqfq}). Furthermore, the axis $T=L$ is a fixed axis.
\label{FixedPoints_determ} 
\end{mytheorem}

\hfill

\begin{mytheorem}
Let $(T^*, L^*)$ be one of the fixed point of the subsystem of equations (\ref{L_tau}) and (\ref{T_tau}) for trust and leverage.
Then, these corresponding equations can be linearised close to 
$(T^*, L^*)$ using a Taylor expansion around the fixed point to yield
{\renewcommand{\arraystretch}{1.5}
\begin{flalign}
\left[
\begin{matrix}
 \frac{\D T}{\D \tau} \\[2ex] \frac{\D L}{\D \tau}
\end{matrix}
\right]
=
{
\left[ 
\begin{matrix}
(2T^*-L^*)(1-T^*)-T^*(T^*-L^*) & -T^*(1-T^*)\\
(1-L^*) \left[ \beta \frac{L_0 - L}{(1-T^*)^2 } + 2 T^* - L^* \right] &-\beta \frac{L_0+T^*-2L^*}{1-T^*} - T^*(1+T^*-2L^*)
\end{matrix} \right]
}
\left[
\begin{matrix}
T-T^* \\[2ex] L-L^*
\end{matrix}
\right], \label{taylor_LT}
\end{flalign}
The 2 by 2 matrix is the Jacobian whose eigenvalues can be studied for the different fixed points. The eigenvalues of the Jacobian evaluated 
for a specific  fixed point indicate whether that fixed point is attractive or repulsive. We find that the point $(T,L)= (1,L_0)$ is attractive while the point $(T,L)= (0,L_0)$ is repulsive.  The axis $T=L$ is (partly)  attractive or repulsive depending on the model parameters $\tilde a, \tilde g, \tilde r$. For $L<L_0$, the axis $T=L$ is  attractive, while it is repulsive for $L>L_0$. This implies that, if $g>r$ (then $L_0>1$), the axis $T=L$ will be entirely attractive for $L\in [0,1]$.}
\label{Theorem_fixedpoints} 
\end{mytheorem}

\hfill

\begin{mytheorem}
The corresponding ROA at the fixed points  $(T,L)= (1,L_0)$ and $(T,L)= (0,L_0)$  is $-\tilde a$. 
The ROA and ROE are equal on the fixed axis $T=L$ and are given by:
\begin{flalign}
&\tilde r_{A} |_{T=L} = \tilde g \frac{1}{1-L} -\tilde r \frac{L}{1-L}.  \label{rE_eq_rA_TL}
\end{flalign}
\label{Theorem_fixedpoints2} 
\end{mytheorem}

\section{Phase portraits of leverage/trust trajectories and associated return on assets} \label{sec:results_fulldynamics}

\subsection{Leverage/trust trajectories and return on assets: three cases} \label{sec:lev_trust_3cases}

\begin{figure}[H]
\centerfloat
\begin{subfigure}{0.6\textwidth}
\includegraphics[trim={0.9cm 0.7cm 0.9cm 0.6cm},clip,width=\textwidth]{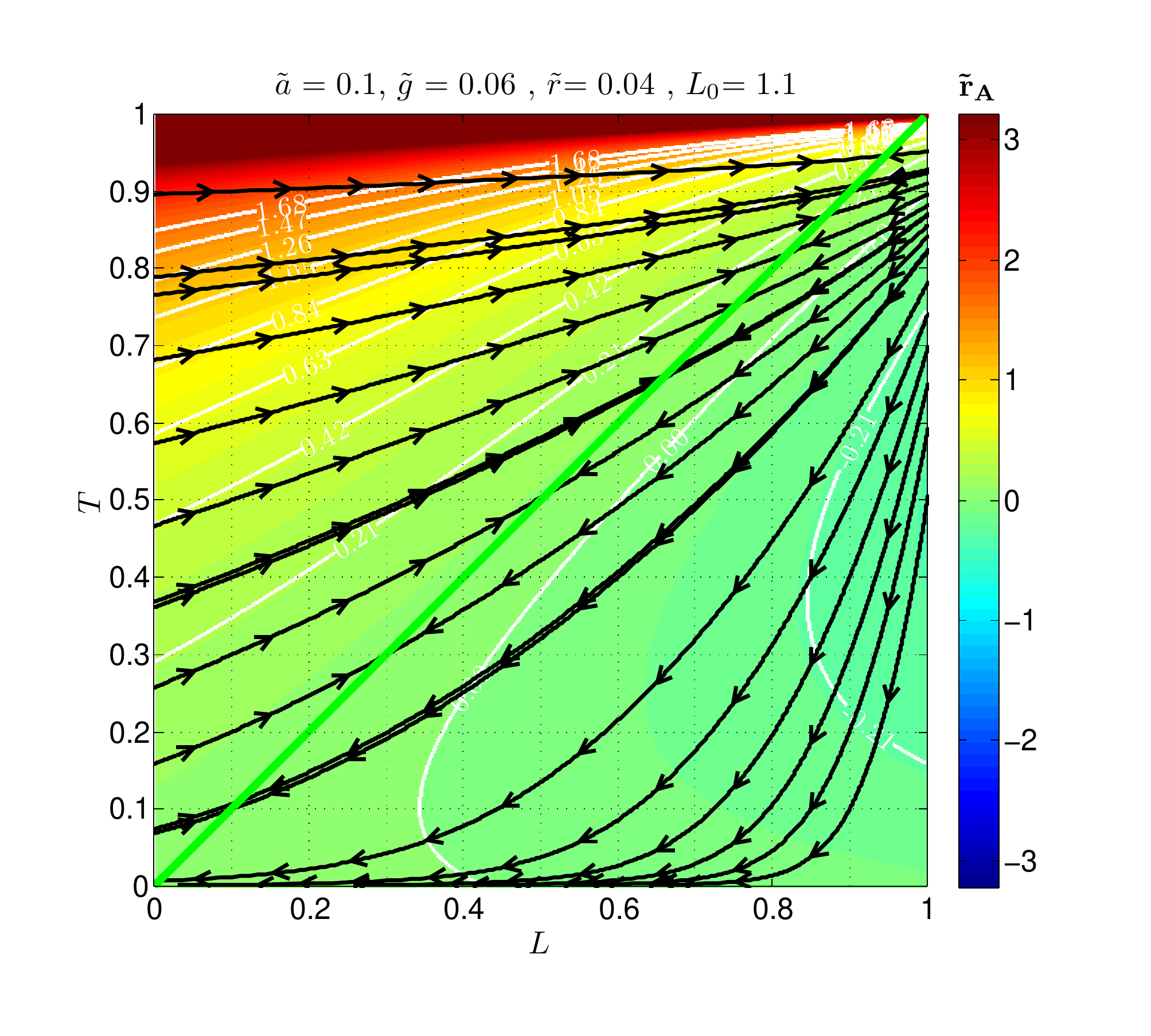} 
     \captionsetup{width=0.84\textwidth}
     \vspace{-0.7cm}
\subcaption{Leverage/trust trajectories (black lines) and ROA ($\tilde r_A$) contour plot.}\label{sf:contour_1}
\end{subfigure}
\begin{subfigure}{0.6\textwidth}
\vspace{0.25cm}
\fboxrule=0pt\relax
\fboxsep=0pt\relax
\noindent\stackinset{r}{15pt}{b}{23.1pt}
  {\color{white}\fbox{\colorbox{white}{\includegraphics[trim={0cm 0cm 0cm 0cm},clip,width=2in]{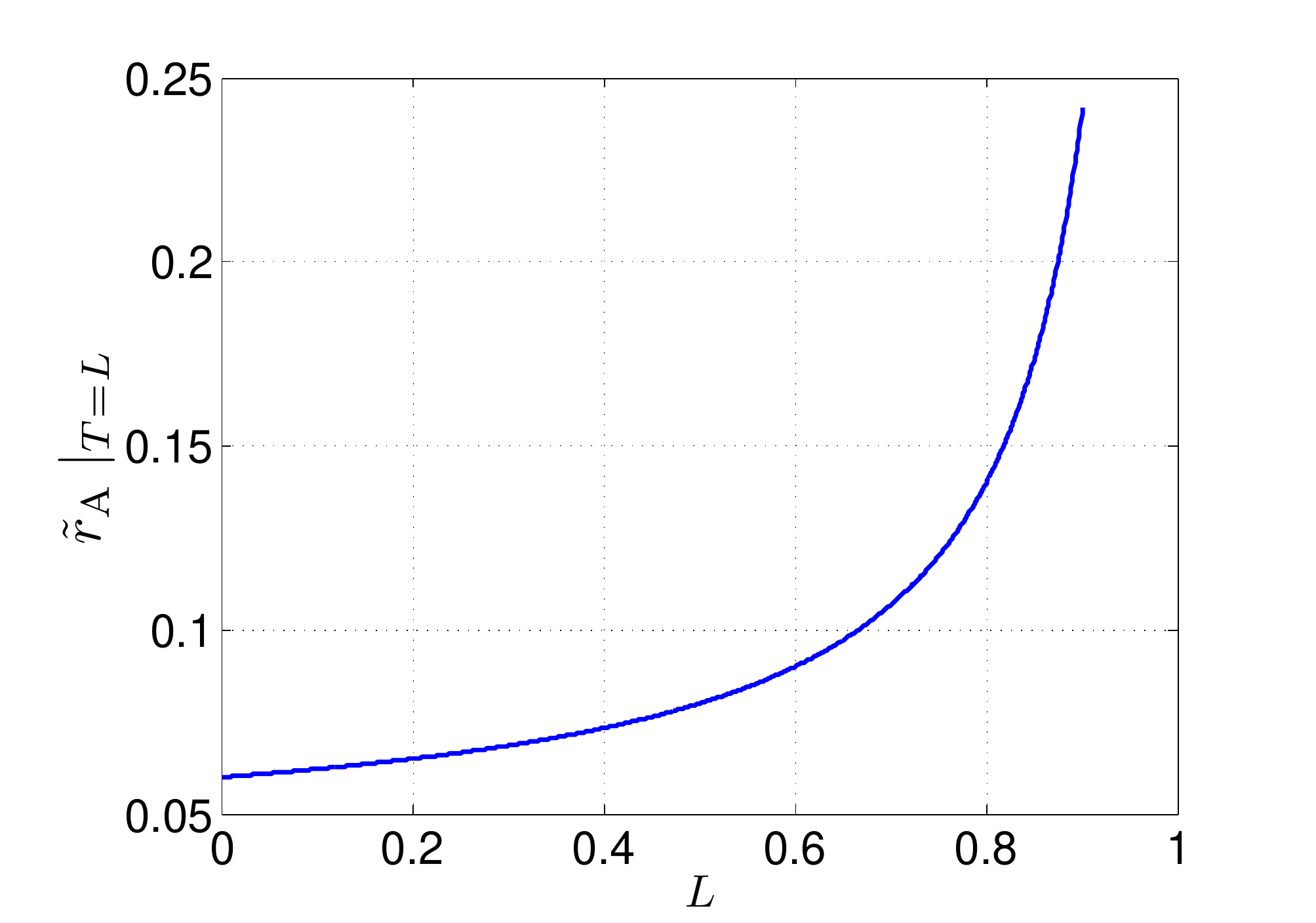}}}}
  {\includegraphics[trim={0.9cm 0.7cm 0.9cm 0.6cm},clip,width=0.94\textwidth]{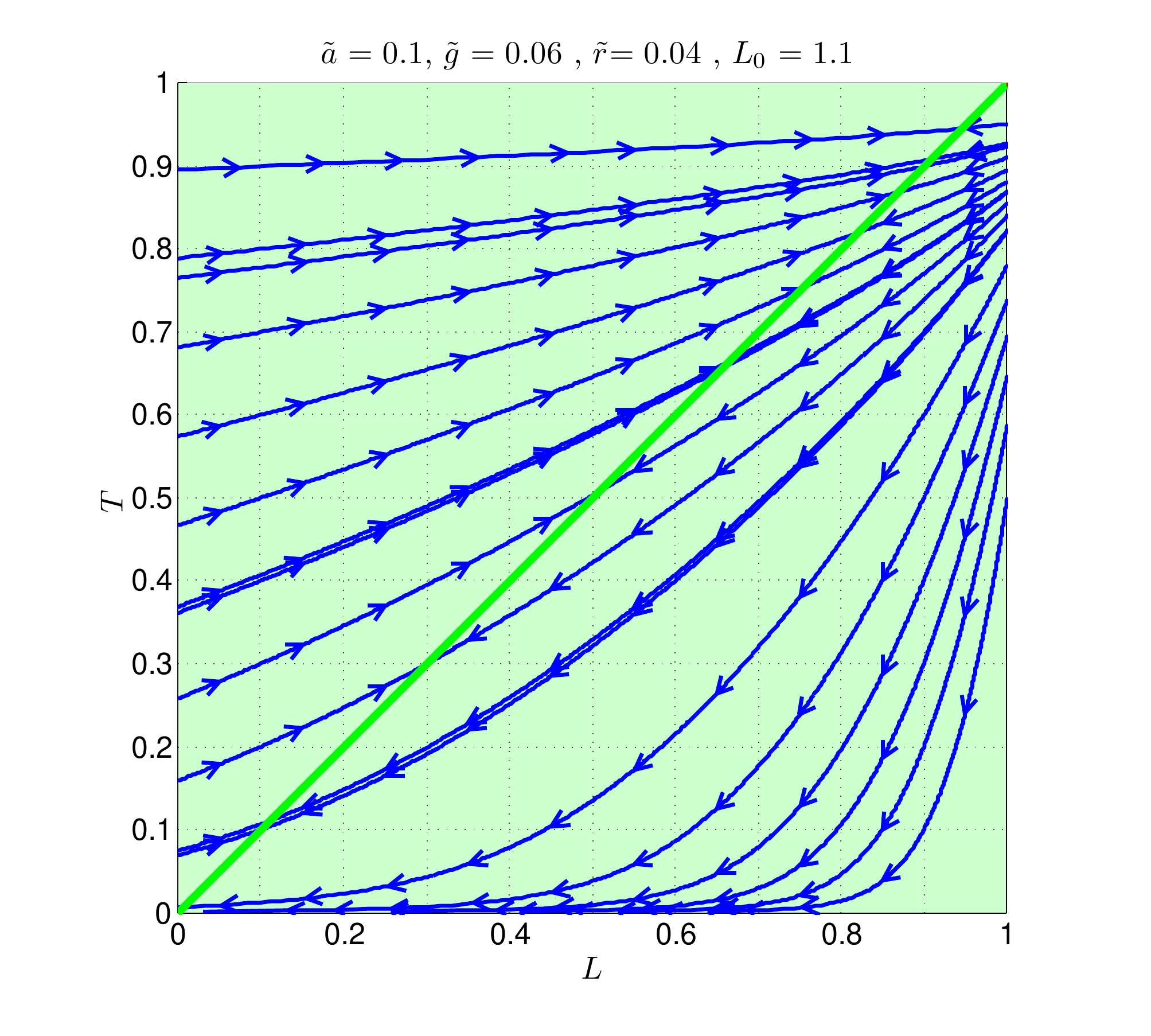}}\medskip
     \captionsetup{width=0.84\textwidth}
          \vspace{-0.2cm}
       \subcaption{Basins of attraction/repulsion and the return on assets on the line $T=L$ as figure inset.}
  \label{sf:inset_1}
  \end{subfigure}
      \vspace{-0.2cm}
  \caption{The case $\tilde g>\tilde r$ is illustrated. It can be observed that all trajectories (black/blue lines) converge to the attractive fixed axis $T=L$ (in green). On the line $T=L$, the return on assets is higher, the higher $L$ is.  Some of the trajectories are highly curved, exemplifying the non-linear dynamics that follows from the model.}
  \label{fig:g_greater_r}
  \end{figure}

\begin{figure}[H]
\vspace{-0.95cm}
\centerfloat
\begin{subfigure}{0.6\textwidth}
\includegraphics[trim={0.9cm 0.7cm 0.9cm 0.6cm},clip,width=\textwidth]{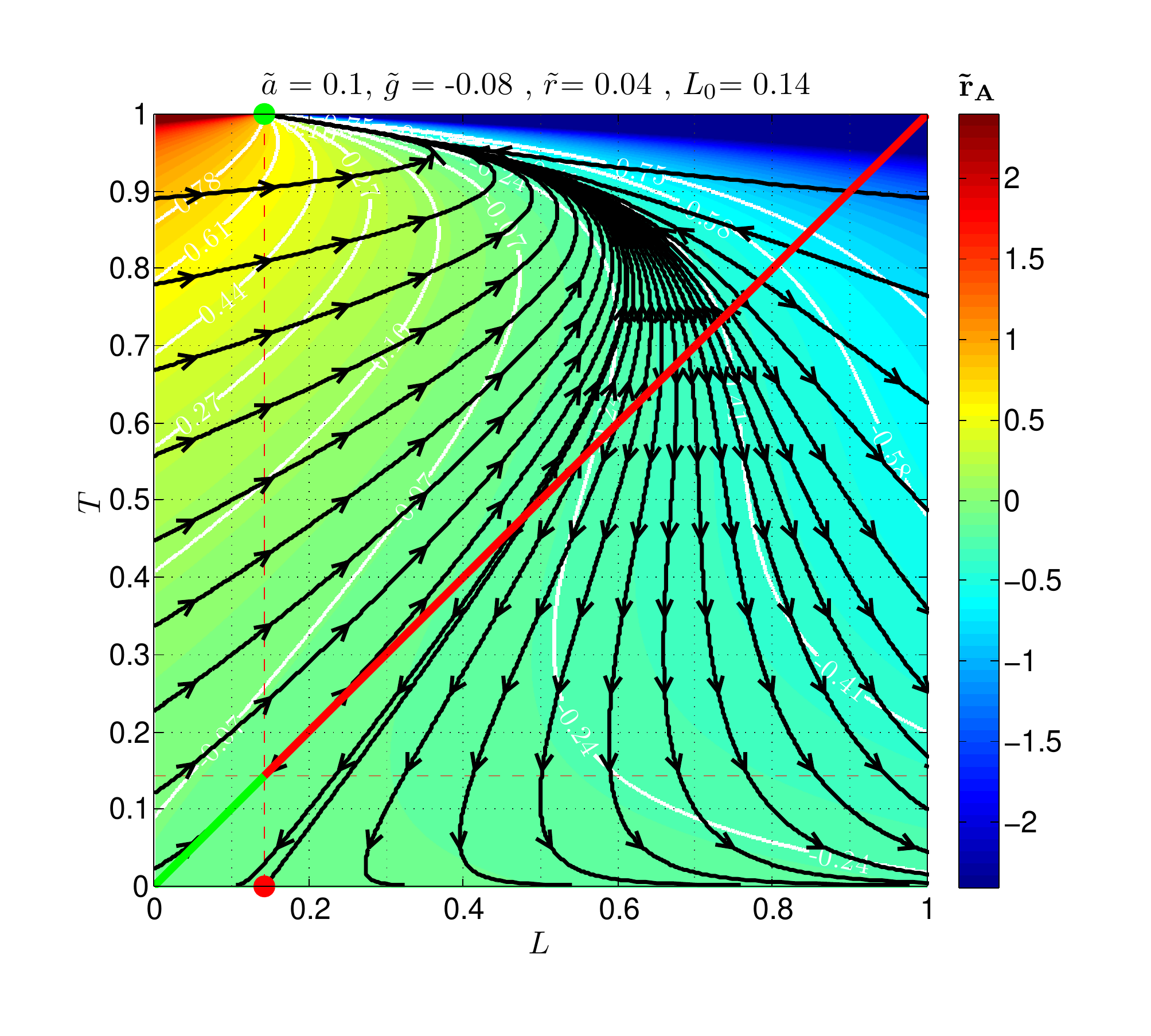}
          \captionsetup{width=0.84\textwidth}
               \vspace{-0.7cm}
\subcaption{Leverage/trust trajectories (black lines) and ROA ($\tilde r_A$) contour plot.}
\label{sf:contour_case2}
\end{subfigure}
\begin{subfigure}{0.6\textwidth}
\vspace{0.25cm}
\fboxrule=0pt\relax
\fboxsep=0pt\relax
\noindent\stackinset{r}{15pt}{b}{23.1pt}
  {\color{white}\fbox{\colorbox{white}{\includegraphics[width=2in]{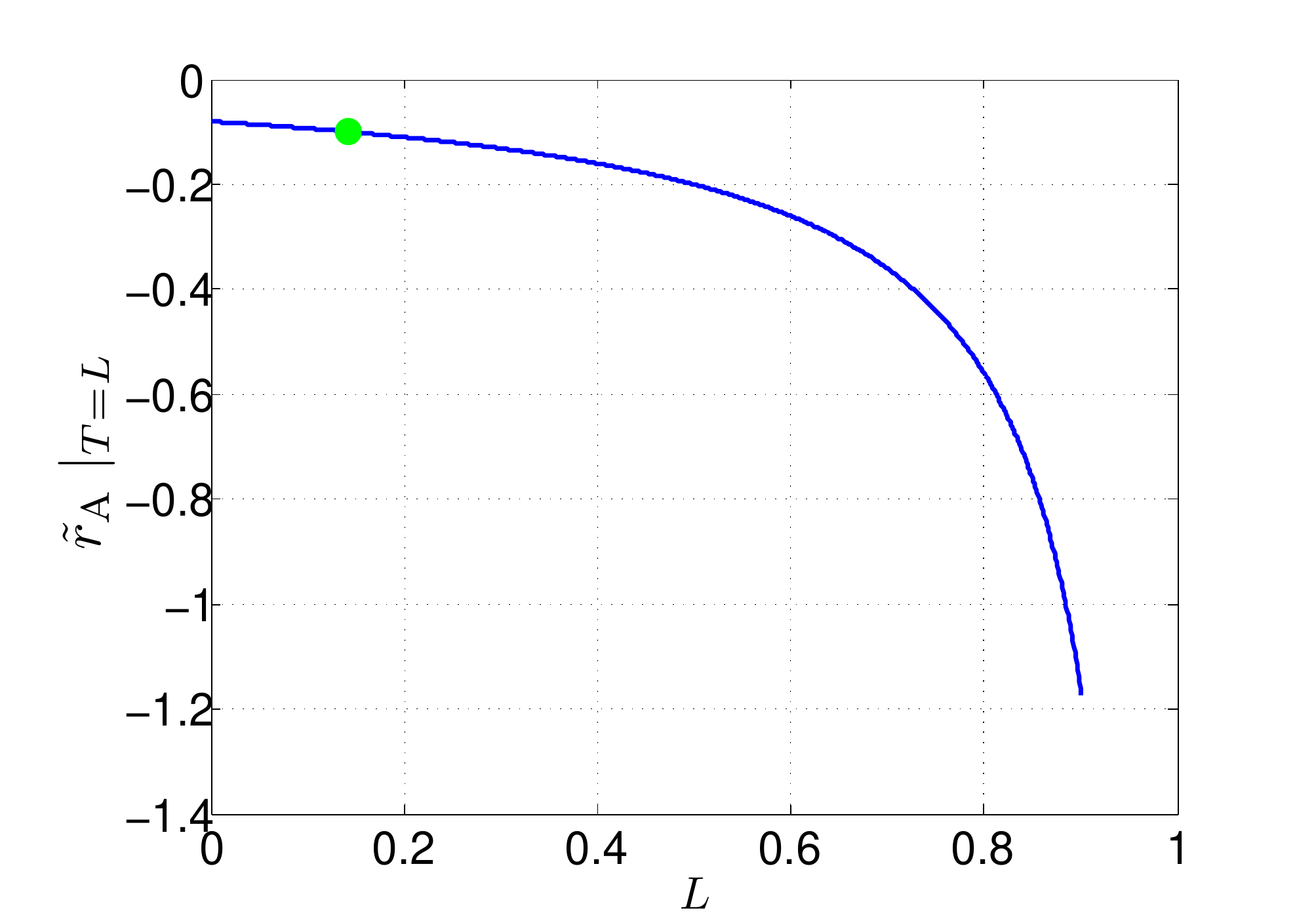}}}}
  {\includegraphics[trim={0.9cm 0.7cm 0.9cm 0.6cm},clip,width=0.94\textwidth]{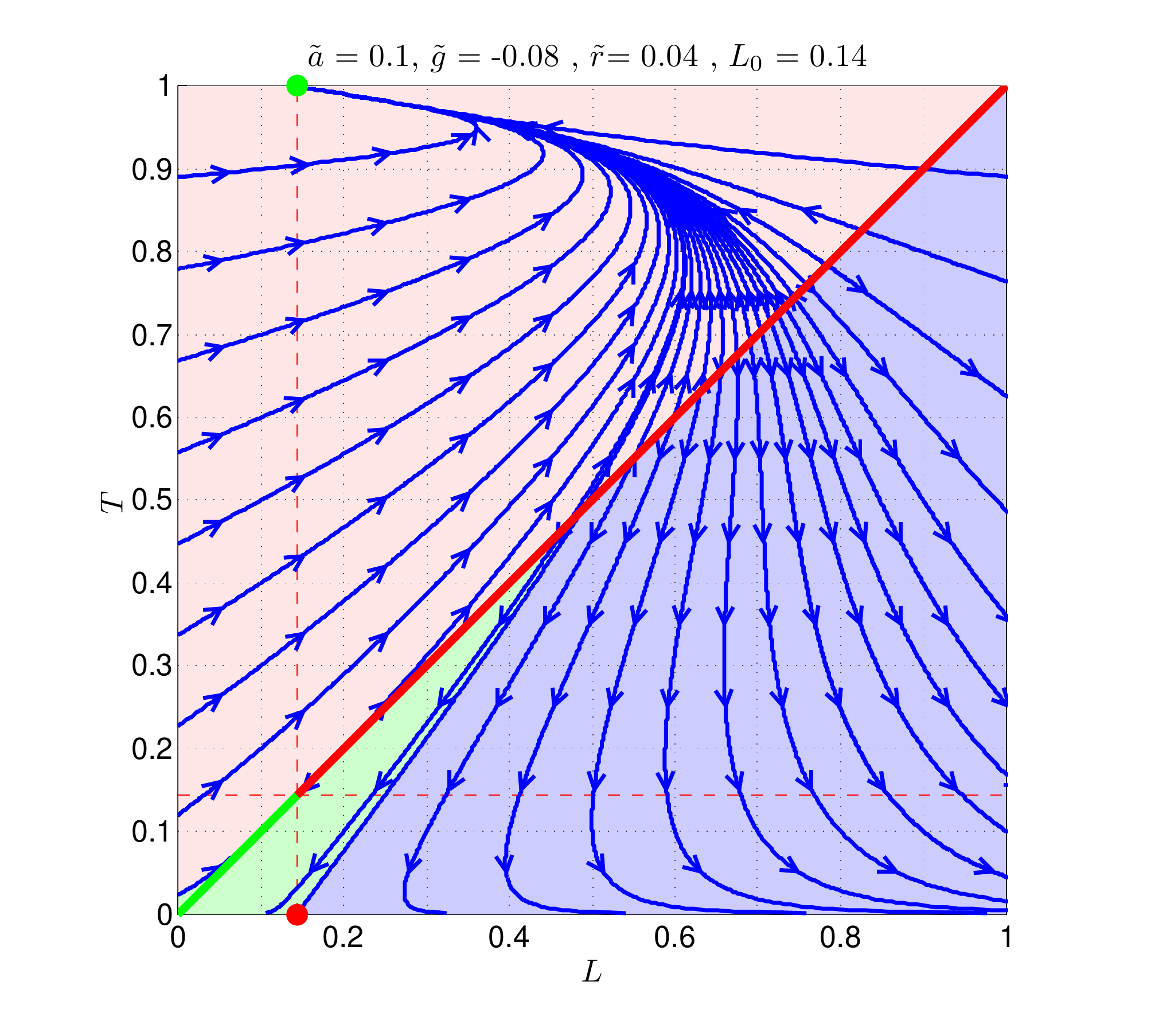}}\medskip
     \captionsetup{width=0.84\textwidth}
          \vspace{-0.2cm}
  \subcaption{Basins of attraction/repulsion and the return on assets on the line $T=L$ and in $(T,L)=(1,L_0)$ as figure inset.}
  \label{g<r_bassins}
  \end{subfigure}
    \vspace{-0.2cm}
  \caption{The case $\tilde g<\tilde r$ with $\tilde g<0$ is illustrated. It can be observed that the  fixed axis $T=L$ is partly attractive (green) and partly repulsive (red). All trajectories in the $T>L$ regime converge to the fixed point $(T,L)=(1,L_0)$, shown as a green dot, where the return on assets is negative ($-\tilde a$). In the $T<L$ regime, most trajectories are expelled from the domain $L\in[0,1]$, while some trajectories converge to the attractive part of the axis $T=L$.}
   \label{fig:g_smaller_r}
  \end{figure}

\begin{figure}[H]
\vspace{-0.7cm}
\centerfloat
\begin{subfigure}{0.6\textwidth}
\includegraphics[trim={0.9cm 0.7cm 0.9cm 0.6cm},clip,width=\textwidth]{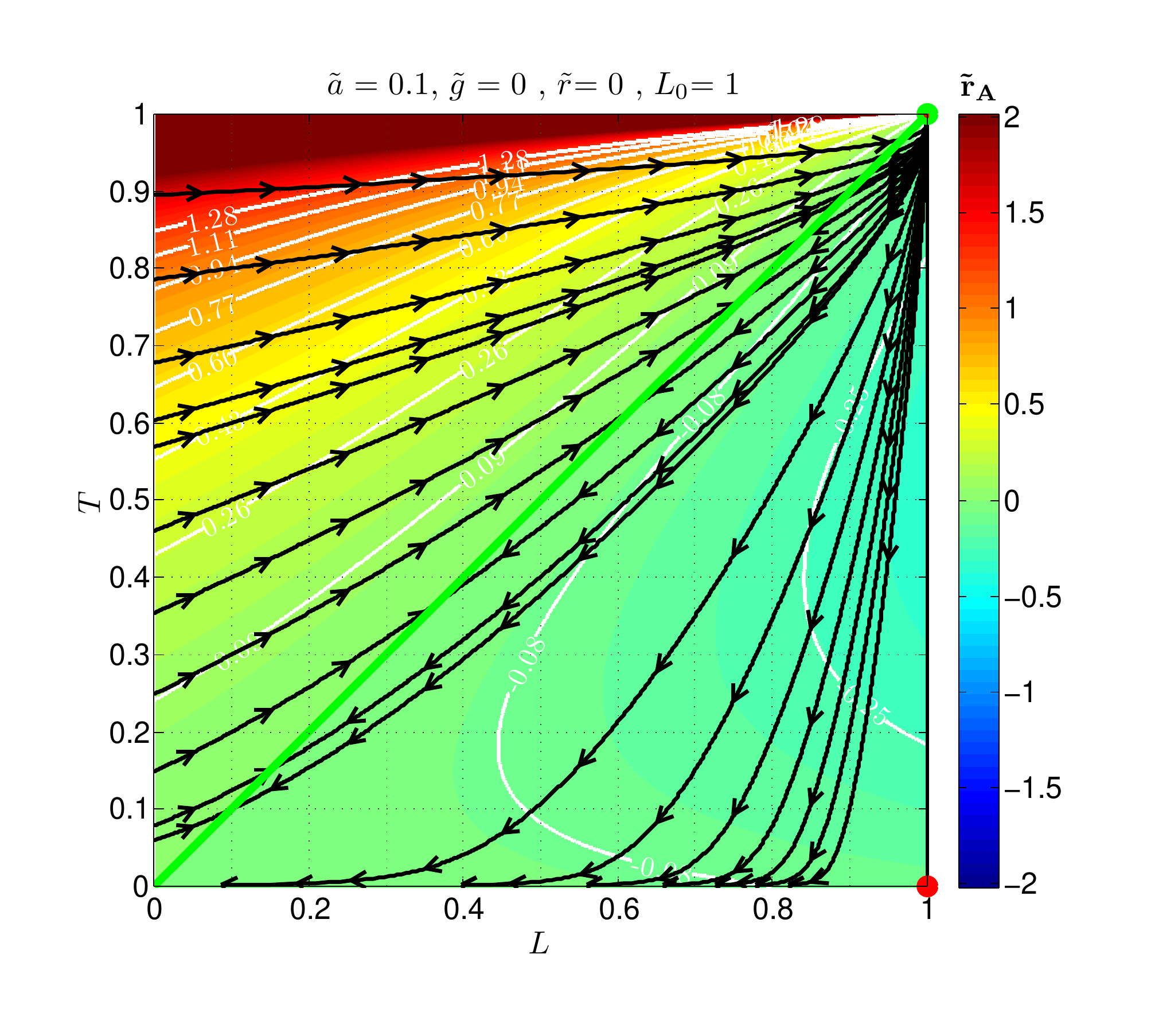} 
     \captionsetup{width=0.84\textwidth}
          \vspace{-0.7cm}
\subcaption{Leverage/trust trajectories (black lines) and ROA ($\tilde r_A$) contour plot.}
\label{sf:contour_3}
\end{subfigure}
\begin{subfigure}{0.6\textwidth}
\vspace{0.25cm}
\fboxrule=0pt\relax
\fboxsep=0pt\relax
\noindent\stackinset{r}{15pt}{b}{23.1pt}
  {\color{white}\fbox{\colorbox{white}{\includegraphics[trim={0cm 0cm 0cm 0cm},width=1.9in]{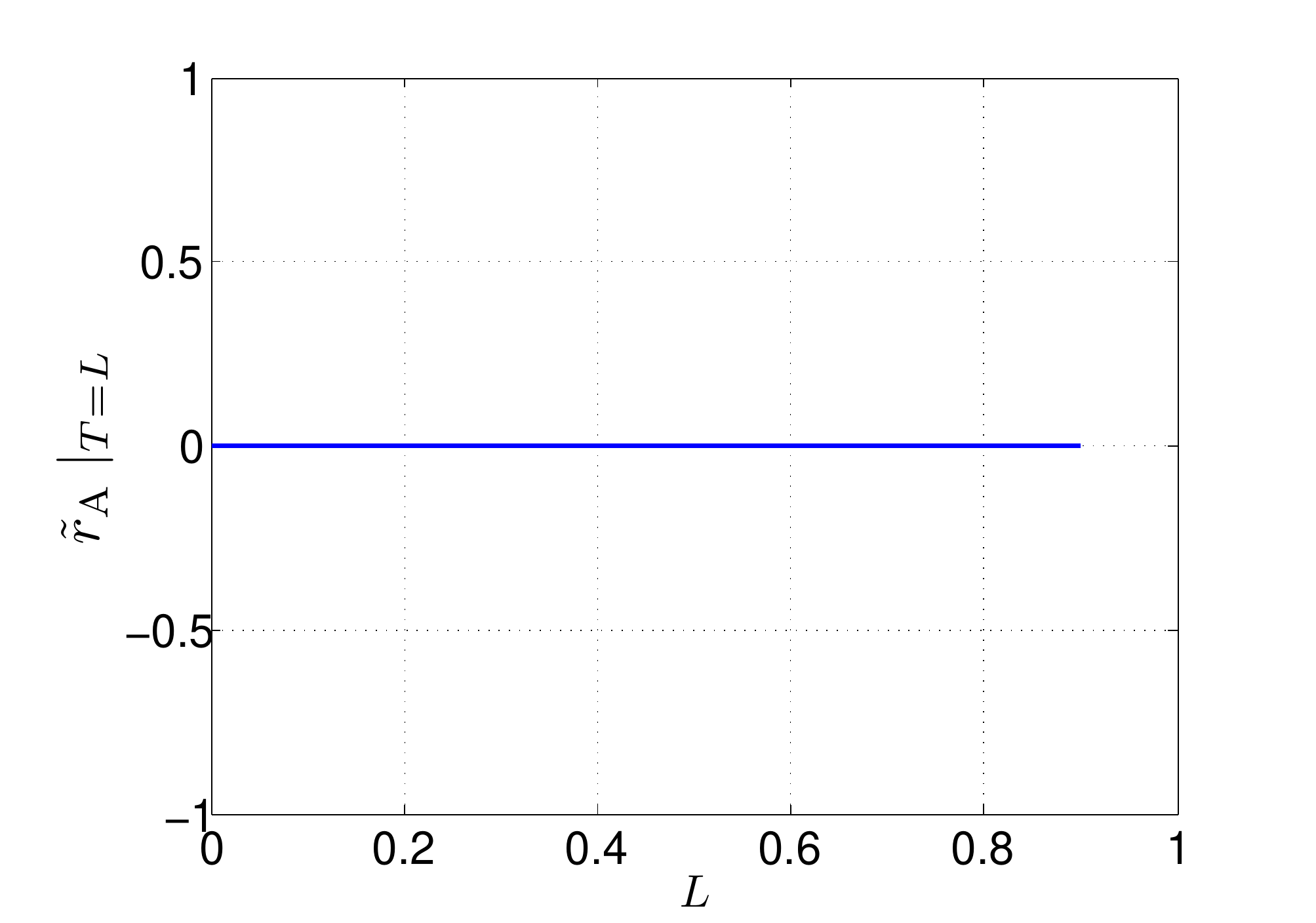}}}}
  {\vspace{0.7cm} \includegraphics[trim={0.9cm 0.7cm 0.9cm 0.6cm},clip,width=0.94\textwidth]{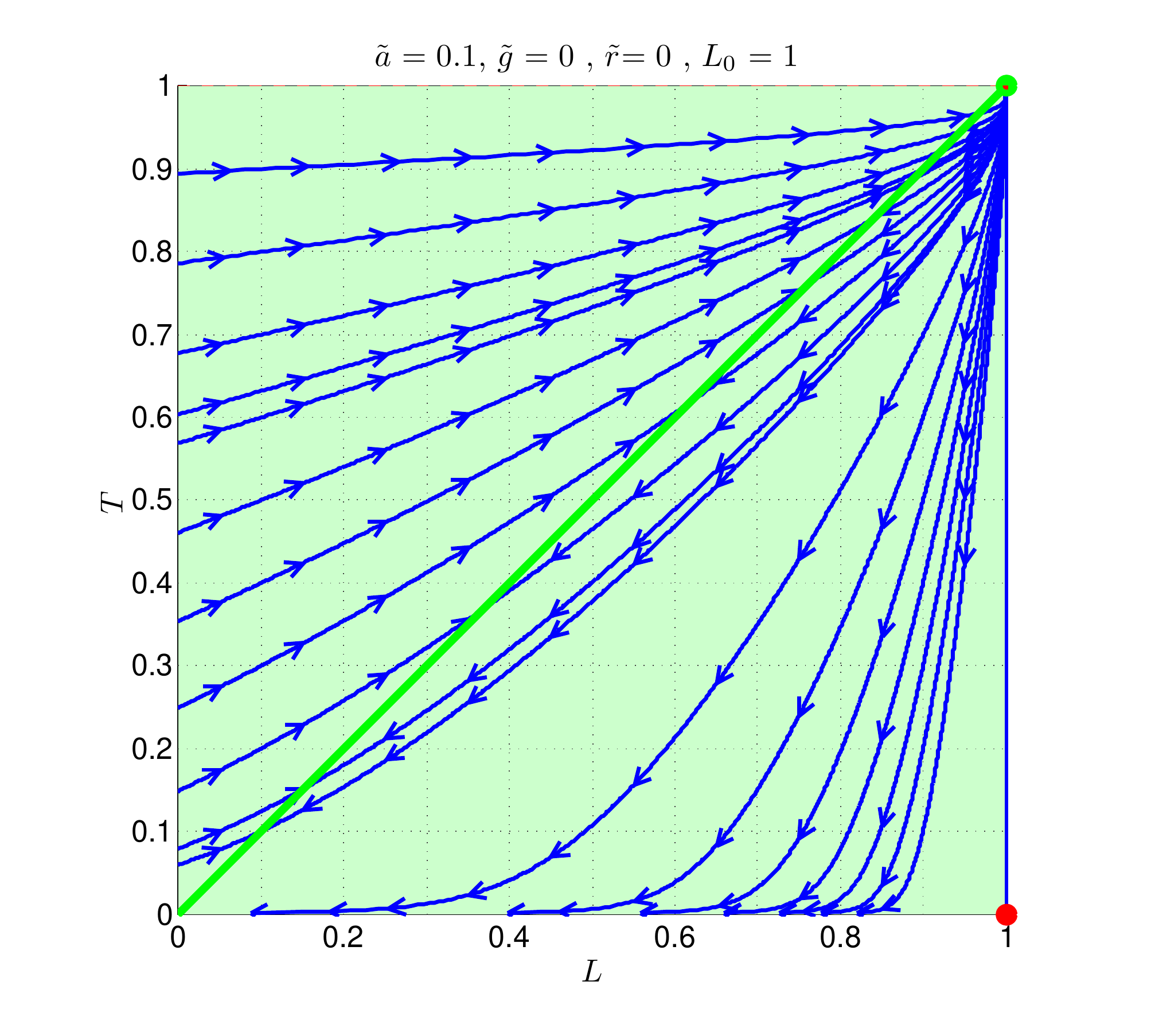}}\medskip
     \captionsetup{width=0.84\textwidth}
          \vspace{-0.2cm}
  \subcaption{Basins of attraction/repulsion and the return on assets on the line $T=L$ as figure inset.}
  \end{subfigure}
  \vspace{-0.2cm}
  \caption{The case $\tilde g=\tilde r=0$ is illustrated. It can be observed that all trajectories (black/blue lines) converge to the attractive fixed axis $T=L$ (in green). On the line $T=L$, the return on assets is equal to zero: the economy is, in the long-term, stuck in a state of no economic growth.}
  \label{fig:g_equal_r}
  \end{figure}

\fig{fig:g_greater_r}-\ref{fig:g_equal_r}  show the dynamics of trust, leverage and return on assets (taken as a proxy for economic growth),
together with their basins of attraction/repulsion, for three regimes: $\tilde g>\tilde r$, $\tilde g< \tilde r$ with $\tilde g<0$, and $\tilde g= \tilde r$.

In each figure, panel (a) shows the non-dimensional return on assets (\eq{r_A_general2}) in color code, corresponding
to the position $(L, T)$ in the diagram. Moreover,  a number of trajectories (representing the vector field) of
leverage and trust given by \eq{LT_final_th} are shown. Green boldface dots and green lines 
indicate attractive fixed points and lines, while the red boldface dots and red lines indicate the unstable points and lines.

In each figure, panel (b) maps the different ``basins of attraction'' of the corresponding attractive
fixed points and lines. The various colours indicate the following:
\begin{enumerate}[topsep=0pt,itemsep=-1ex,partopsep=1ex,parsep=1ex]
\item[$\bullet$] Light green area:  domain of attraction towards the attractive axis $T=L$.
 \item[$\bullet$] Light red area: domain of attraction towards the fixed point $(T, L)=(1,L_0)$. 
 \item[$\bullet$] Light blue area: the leverage increases (it may locally decrease) so that the trajectories eventually leave the 
domain $L\in [0,1]$. Mathematically, we have $\tilde r_L>0$ (locally $\tilde r_L\leq0$ may occur). 
  \end{enumerate}
 
The insets in panels (b) of \fig{fig:g_greater_r}-\ref{fig:g_equal_r} 
show the return on assets on the axis $T=L$.  The return on assets increases, decreases or stays constant when $L$ increases, depending on which case is studied ($\tilde g>\tilde r$, $\tilde g<\tilde r$, or $\tilde g=\tilde r$).
Note that the return on assets is equal to the return on equity on the axis $T=L$ (Theorem \ref{Theorem_fixedpoints2}).

\clearpage

\subsection{The mechanism of reward and penalty in terms of long-term return on assets}

An interesting feature of \fig{fig:g_greater_r} is that the leverage/trust trajectories are upward sloping in the $T>L$ domain, while they are downward sloping in the $T<L$ domain. The higher $T_{\text{stationary}}=L_{\text{stationary}}$, the higher the stationary ROA is. Hence, regular economies ($\tilde g>\tilde r$) that first establish trust and then increase leverage are rewarded with the highest steady-state (long-term) return on assets, while economies that significantly increase leverage before having established trust are penalized with a lower long-term return on assets. Since $\tilde r_A |_{T=L} \propto \frac{1}{1-L}$ the reward for the patient economy that first builds trust is substantial. 

This existence of a reward when trust growth precedes leverage growth (and $T>L$) and the presence of penalty when $T<L$ are not always prevalent.
In the case where $\tilde g<\tilde r$  with $\tilde g$ negative (see \fig{fig:g_smaller_r}), it can be observed that, in economies where trust is  established first ($T>L$), the trajectories eventually reach a steady-state point $(T,L)=(1, L_0)$ where the return on assets is negative ($\tilde r_A|_{(T=1, L=L_0)} =-\tilde a$). Some economies in the $T<L$ regime are better off, namely those in the light green area in \fig{g<r_bassins} that converge to the attractive part of the axis $T=L$ where the ROA is still negative but is larger than $-\tilde a$.
The other economies in the $T<L$ regime (light blue area in \fig{g<r_bassins}) are, however, in a very detrimental regime.
The  trajectories  there  will eventually be expelled from the domain $L \in [0,1]$. Furthermore, 
the trust will be rapidly destroyed  and will reach values close to zero, which implies that virtually no credit is available in the economy (if the trust is 0, there are no assets that can qualify as collateral to obtain a loan). On top of that, as can be observed from the contour plot of \fig{sf:contour_case2}, the return on assets is negative in the light blue area.

In the special case where $\tilde g=\tilde r=0$ (\fig{fig:g_equal_r}), there is no reward or penalty whatsoever, since all trajectories move to the steady-state line $T=L$ where the return on assets is zero.   
 This steady-state is reminiscent of the economic state in Japan since 1990 (often referred to as the ``two lost decades'') and in the Eurozone since the US-based subprime crisis in 2008 and the sovereign debt crisis in Europe starting in 2010. Japan and the Eurozone have
 been characterised by essentially vanishing interest rates and very low economic growth. In  January 2015, the European Central Bank (ECB) announced that it will start to buy \euro{60}bn of bonds each month \citep{ECB} from March 2015 for a determined period; this quantitative easing (QE) programme is a measure meant to revitalise the economy of the Eurozone. In the context of our model in terms 
 of interacting assets, leverage and trust, this QE policy aims at boosting the parameter $\tilde g$, which 
 ensures that the long-term ROA (on the axis $T=L$) increases. We will analyse the impact of negative shocks
 and policy response within the context of our model in section  \ref{sec:leverage_trust_regime_shifts}.

\subsection{Transient costs before convergence to the beneficial fixed points for $\tilde g>\tilde r$}

In all cases, by definition of what is a fixed point, the steady-state growth of leverage is zero (i.e. $L_{\text{stationary}}=L^*=\text{constant}$), or equivalently (based on \eq{debt_lev_assets_growth}) the asset and debt value grow or shrink at the same pace in the stationary (long-term) situation. 

In the short-run however, the leverage is non constant and its dynamics may lead to non-intuitive effects.
For example, for $\tilde g>\tilde r$ (\fig{fig:g_greater_r}), in the $T>L$ regime,  the leverage increases in the short-run. The long-term beneficial  state of high ROA thus comes at a transient cost ($\tilde r_L>0 \; \Leftrightarrow \; \tilde r_D>\tilde r_A$ in the short-run).
The transient cost is  the result of a convergence of the economy to a sustainable stationary fixed point (where the ROA is positive) with maximum output ($L_{\text{stationary}}=T_{\text{stationary}}$ with a value close to 1). The leverage increases in the short-run, which leads to a beneficial stationary state with a high ROA; this makes the transient fast growth of debt tolerable in view of the beneficial long-term goal.

\section{Leverage/trust trajectories with regime shifts} \label{sec:leverage_trust_regime_shifts}

This section addresses the question of when and how should a central bank intervene in the face of a shock to an economy.
Should a central bank directly act or should it best be patient and intervene  relatively late? 
In the context of our model, we explore these questions by studying the trajectories in the leverage/trust phase portrait
under different regime shifts of the model parameters $\tilde g$ and $\tilde r$. 

The model parameters $\tilde r$ and $\tilde g$  can be targeted by the central bank using  conventional policy tools (open market operations, standing facilities, minimum reserve requirements) or unconventional tools (quantitative easing). 
For instance, a standard monetary intervention of a central bank at a time of crisis is to target a decrease of the (short-term) interest rate.
According to general economic theory, a lower interest rate results in lower cost of borrowing, making it more attractive.
The goal is to boost the economy by enticing economic agents to take more risks by investing in potential sources of economic growth.
A short-term effect is also to encourage consumption, which has an immediate effect of GDP (see e.g. \citet{Dalio}).

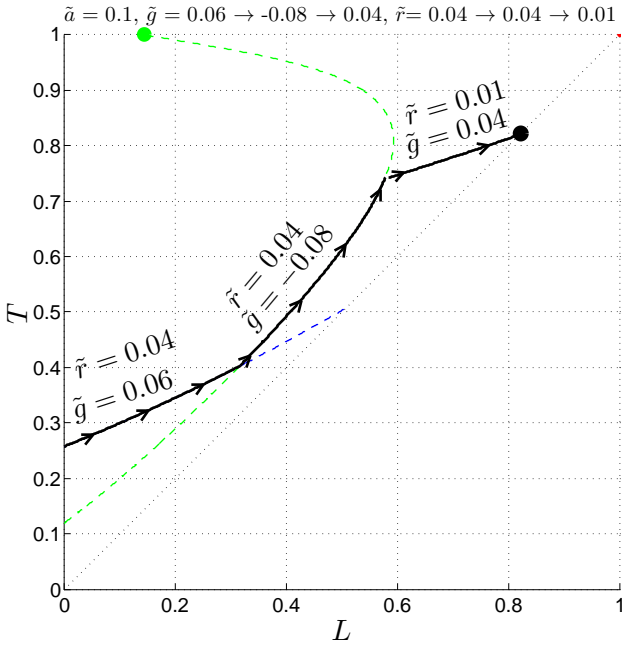
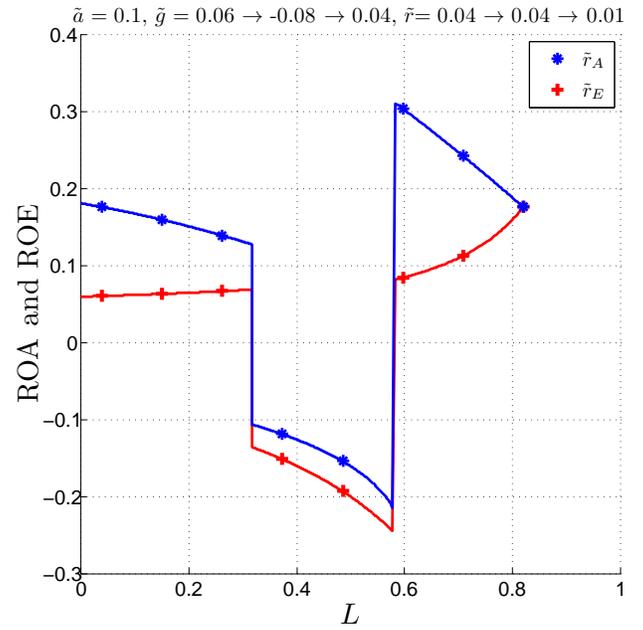
\begin{figure}[H]
\vspace{-0.6cm}
\centerfloat
\begin{subfigure}{0.6\textwidth}
\begin{picture}(0,0)
\put(2,-240){\includegraphics[trim={0 0.73cm 0 0.6cm},clip,width=\textwidth]{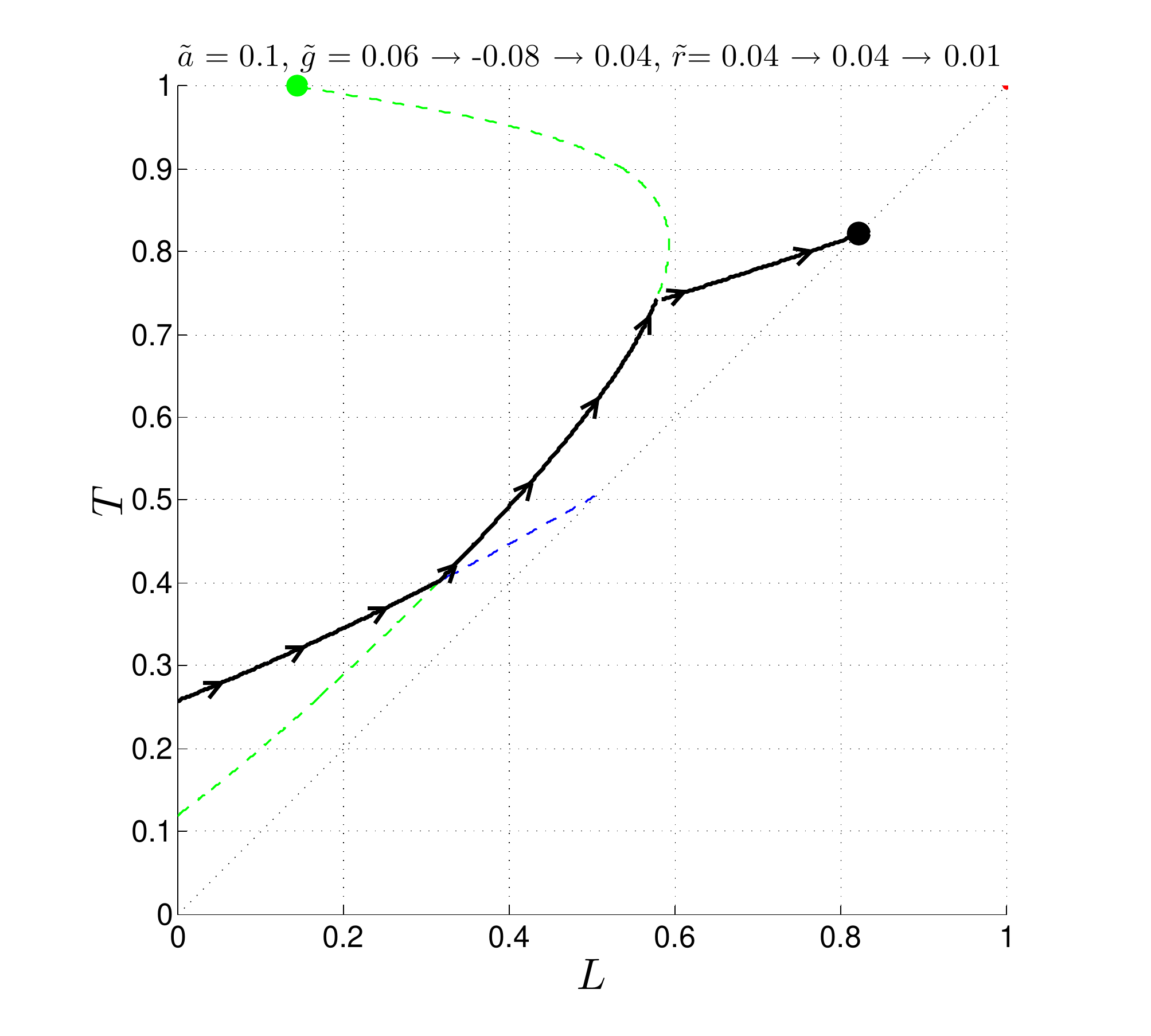}}
\put(175,-55){\rotatebox{16}{$\tilde g = 0.04$}}
\put(175,-44){\rotatebox{16}{$\tilde r = 0.01$}}
\put(113,-122){\rotatebox{50}{$\tilde g = -0.08$}}
\put(107,-114){\rotatebox{50}{$\tilde r = 0.04$}}
\put(49,-155){\rotatebox{18}{$\tilde g = 0.06$}}
\put(50,-140){\rotatebox{18}{$\tilde r = 0.04$} }
\end{picture}
\vspace{8.5cm}
   \captionsetup{width=0.84\textwidth}
\subcaption{Leverage/trust trajectory (black line), where the dashed lines indicate the continuation of the trajectory in case no regime shift would have been imposed.}
\label{sf:medium_path}
\end{subfigure}
\begin{subfigure}{0.6\textwidth}
\includegraphics[trim={0 0.73cm 0 0.6cm},clip,width=\textwidth]{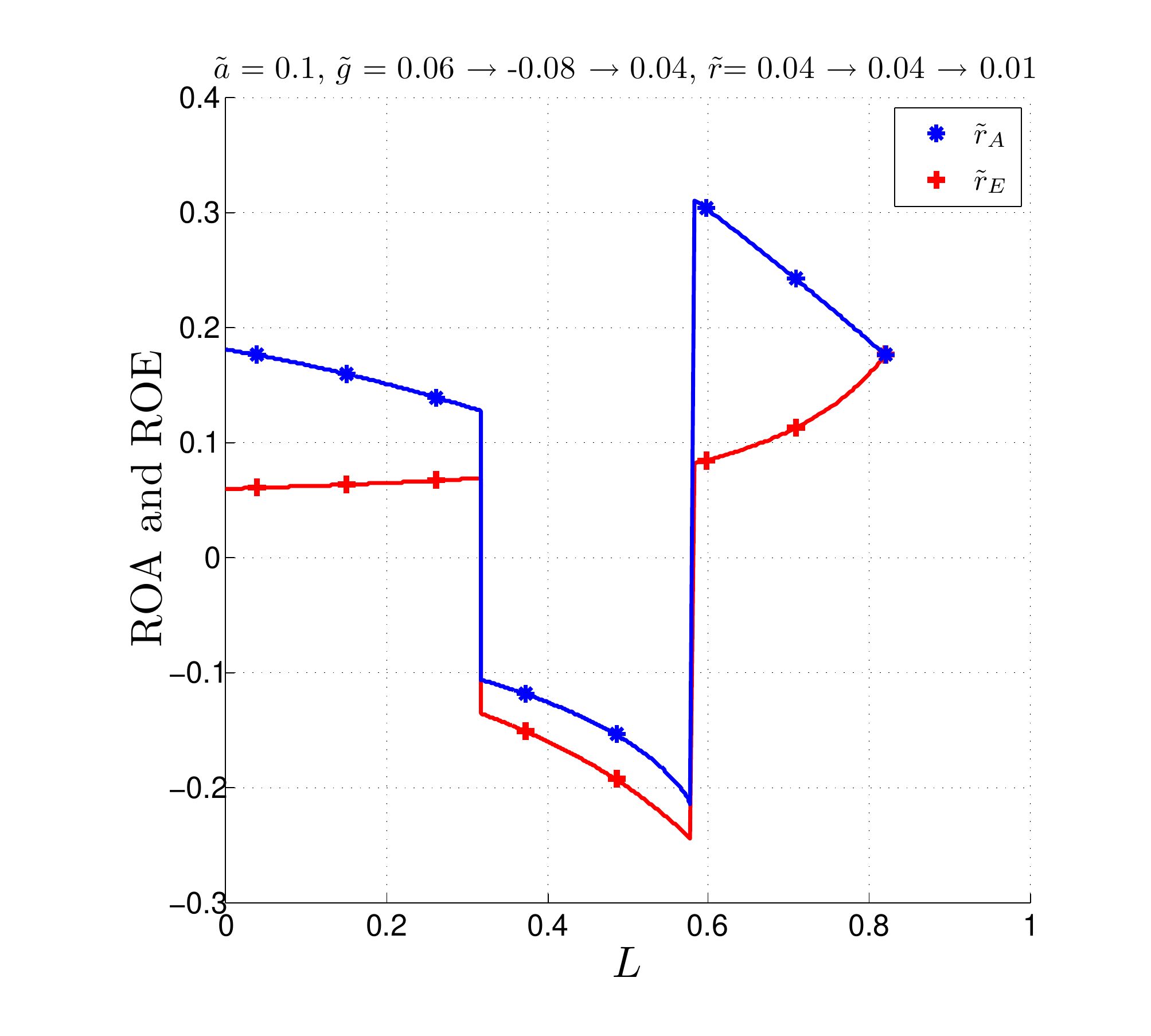}
\vspace{-0.4cm}
   \captionsetup{width=0.84\textwidth}
\subcaption{This graph shows the ROA (in blue) and the ROE (in red) as a function of leverage. The ROA and ROE are endogenous variables. }
\label{sf_medium_quantities_rEetc}
\end{subfigure}
\vspace{-.2cm}
\caption{Leverage/trust trajectories (panel (a)) and their corresponding ROA and ROE (panel (b)) in the presence of regime shifts imposed in the model parameters $\tilde g$ and $\tilde r$. Model parameters are given above the plots. The arrows show the evolution during each regime.  The regime shifts occur at the kinks in the leverage/trust trajectory. They correspond to the jumps in the ROA, ROE plot in panel (b). The first regime shift at constant interest rate $\tilde r$
and with a strong drop of $\tilde g$ is supposed to represent the shock of a financial and/or economic crisis. The second regime shift 
represents the intervention of the central bank in the model economy, which decreases the target interest rate and boosts the
EBITA/Assets ratio. }
\label{fig:medium}
\end{figure}

\begin{figure}[H]
\vspace{-0.5cm}
\centerfloat
\begin{subfigure}{0.6\textwidth}
\includegraphics[trim={0 .73cm 0 0.6cm},clip,width=\textwidth]{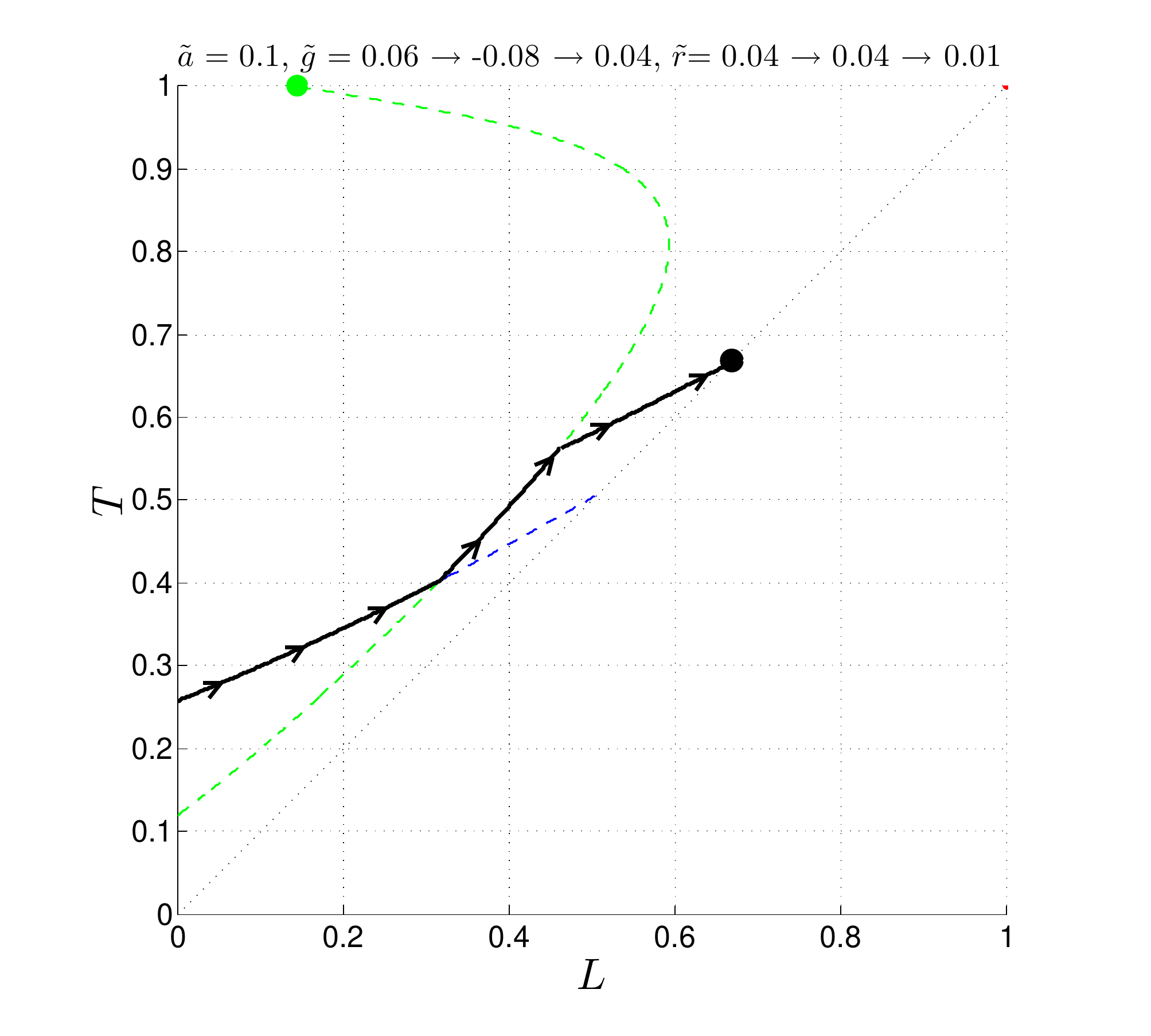} 
   \captionsetup{width=0.84\textwidth}
   \vspace{-.5cm}
\subcaption{Leverage/trust trajectory (black line), where the dashed lines indicate the continuation of the trajectory in case no regime shift would have been imposed.}
\label{sf:early_path}
\end{subfigure}
\begin{subfigure}{0.6\textwidth}
\includegraphics[trim={0 .73cm 0 0.6cm},clip,width=\textwidth]{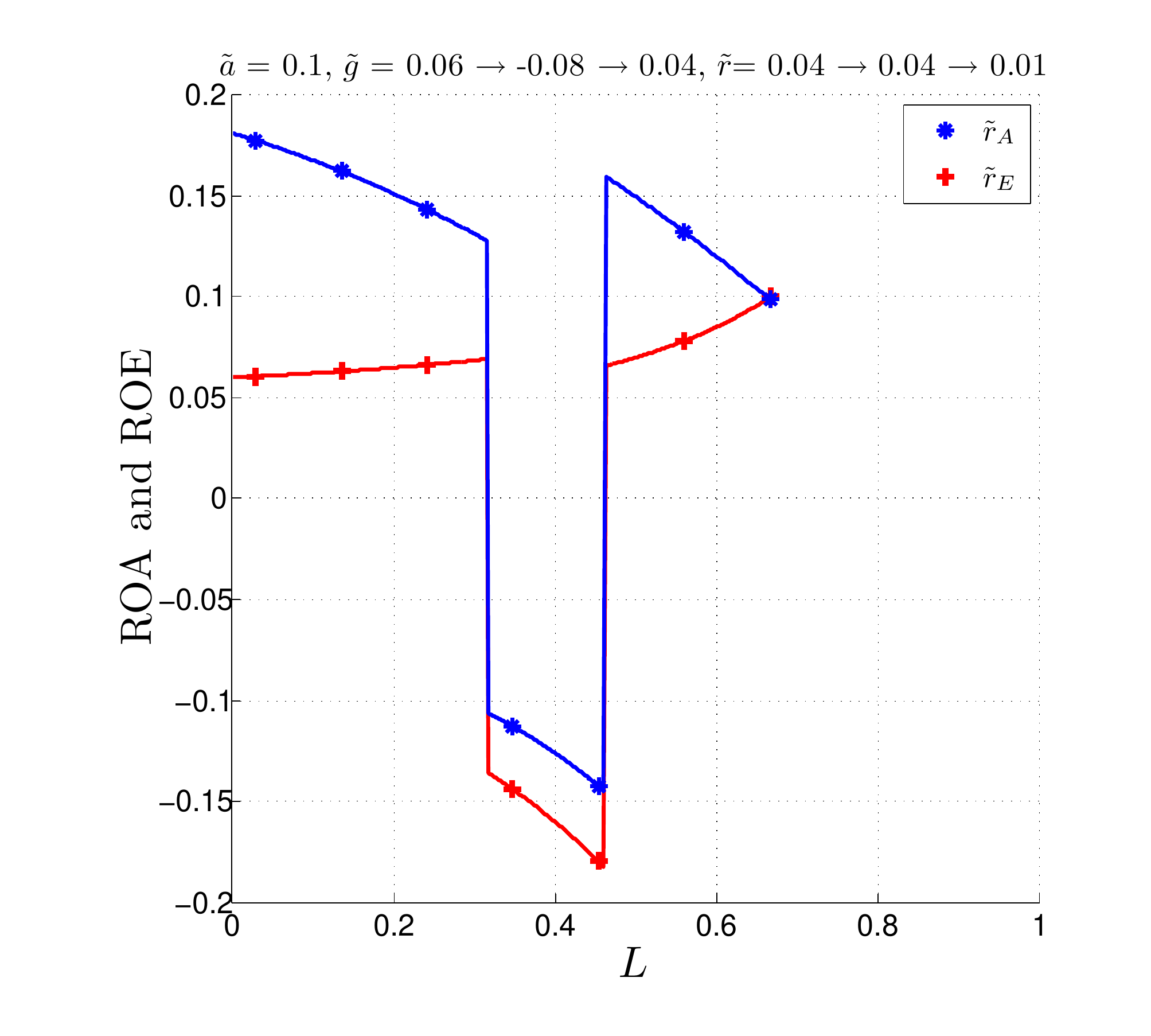}
   \captionsetup{width=0.84\textwidth}
      \vspace{-.5cm}
\subcaption{This graph shows the ROA (in blue) and the ROE (in red) as a function of leverage. The ROA and ROE are endogenous variables.}
\end{subfigure}
\vspace{-0.2cm}
\caption{Same as figure \ref{fig:medium} but with an earlier intervention of the central bank, represented by a quicker second regime shift.}
\label{fig:early}
\end{figure}

\begin{figure}[H]
\vspace{-0.6cm}
\centerfloat
\begin{subfigure}{0.6\textwidth}
\includegraphics[trim={0 0.73cm 0 0.6cm},clip,width=\textwidth]{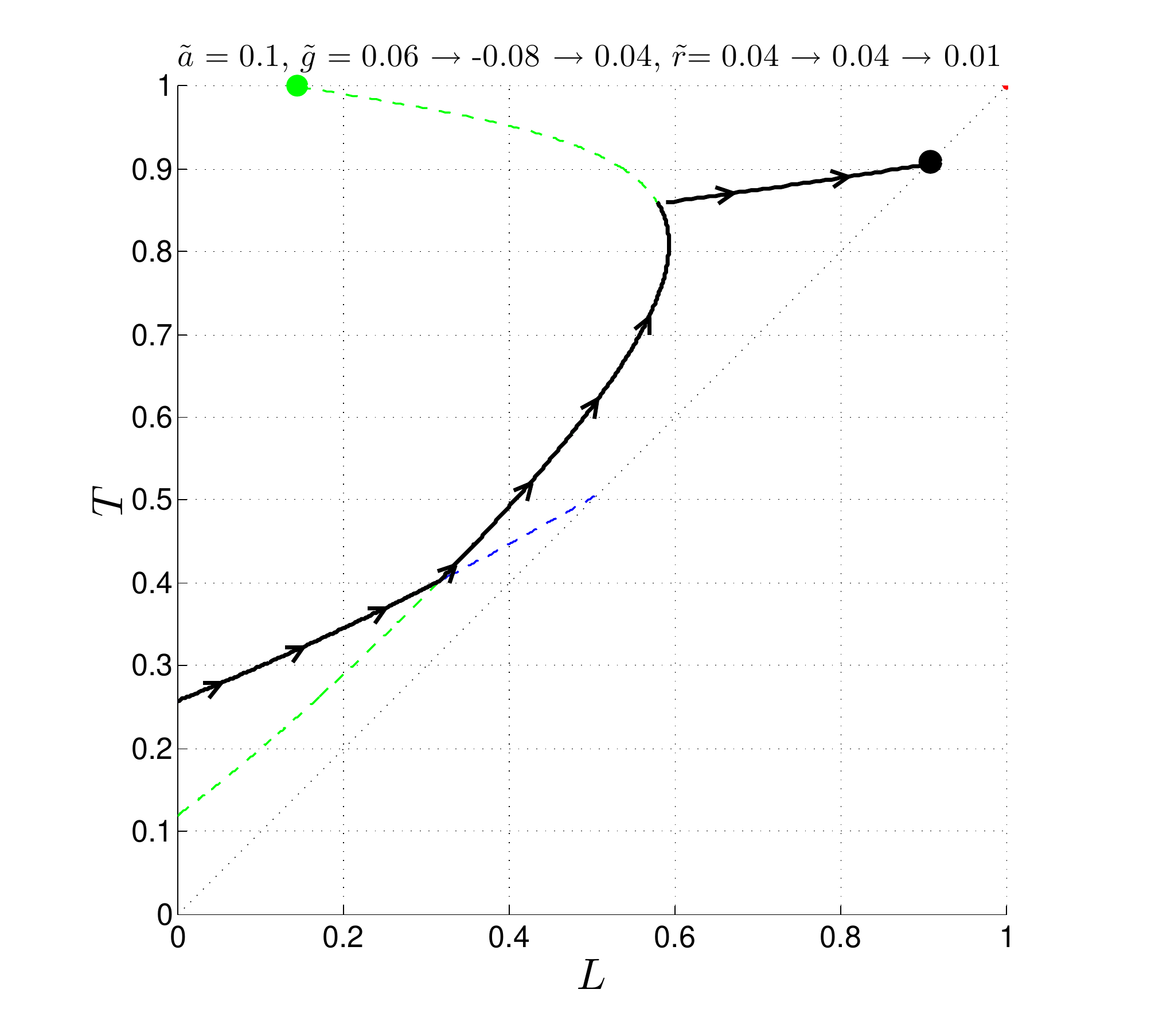} 
   \captionsetup{width=0.84\textwidth}
      \vspace{-.5cm}
\subcaption{Leverage/trust trajectory (black line), where the dashed lines indicate the continuation of the trajectory in case no regime shift would have been imposed.}
\label{sf:late_path}
\end{subfigure}
\begin{subfigure}{0.6\textwidth}
\includegraphics[trim={0 0.73cm 0 0.6cm},clip,width=\textwidth]{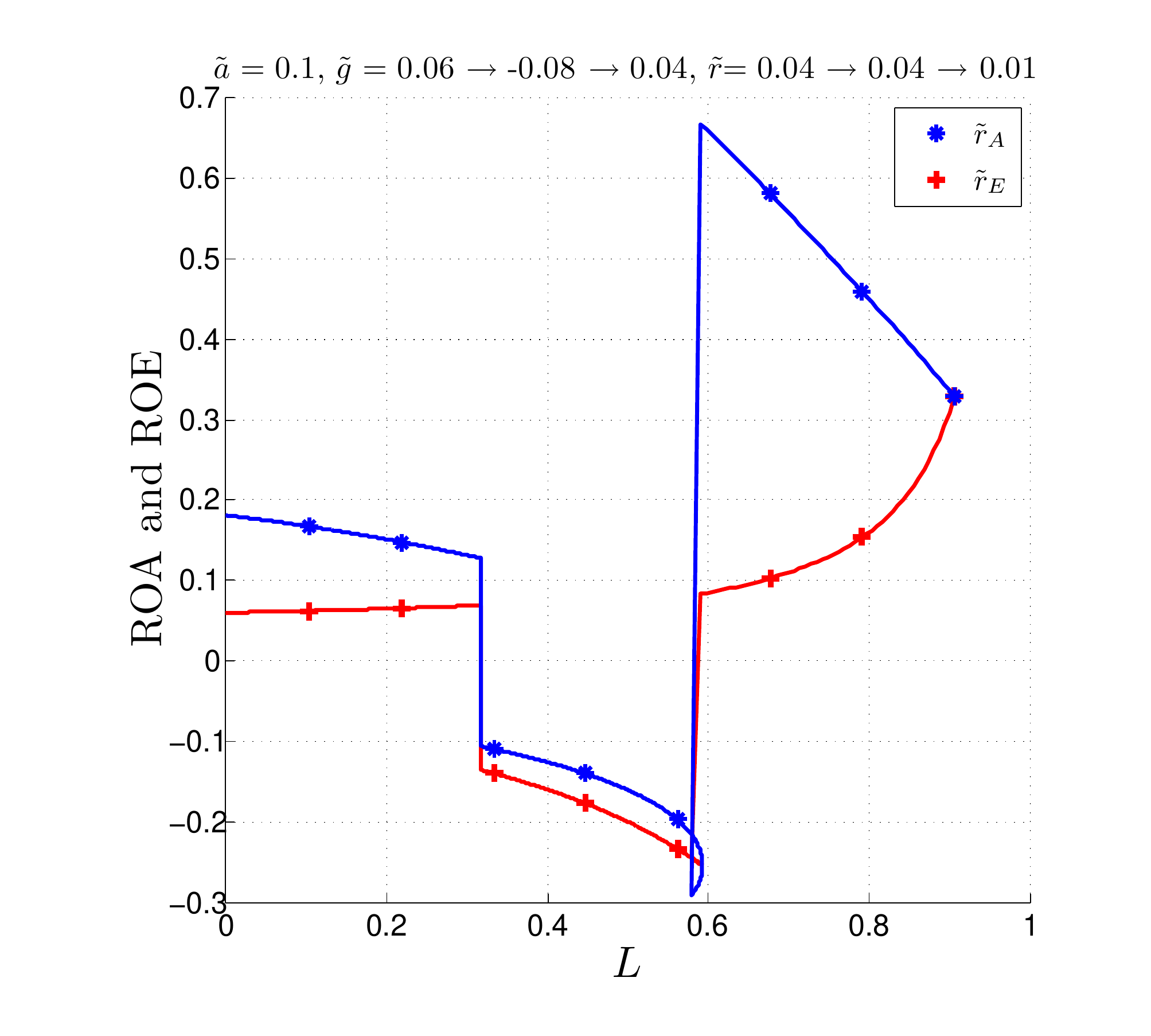}
   \captionsetup{width=0.84\textwidth}
      \vspace{-.5cm}
\subcaption{This graph shows the ROA (in blue) and the ROE (in red) as a function of leverage. The ROA and ROE are endogenous variables.}
\end{subfigure}
\caption{Same as figure \ref{fig:medium} but with a later intervention of the central bank, represented by a delayed second regime shift.}
\label{fig:late}
\end{figure}

\fig{fig:medium}-\ref{fig:late} depict single leverage/trust trajectories and their corresponding ROA and ROE.  
Regime shifts in the model parameters $\tilde g$ and $\tilde r$ are introduced along the trajectory, which 
flows in the direction given by the arrows. 
The relevant model parameters are given above each plot.
The regime shifts in the exogenous model parameters $\tilde g$ and $\tilde r$ attempt to capture
the main aspects of economic reality. For example, the EBITA/Assets ratio ($\tilde g$) of firms might abruptly change due to market changes or as a result of being (indirectly) targeted through a central bank's policy. The interest rate ($\tilde r$) can also abruptly change as a result of a  central bank's policy to change its  interest rate target.

Panels (a) depict the leverage/trust trajectories (in black), with the initial condition $(T,L) = (0.26,0)$.  Regime shifts in $\tilde g$ and $\tilde r$ are imposed along the trajectory. 
The black dot indicates the steady-state point associated with the last values of the parameters. The dashed coloured lines show the continuation of trajectories if no regime shift would have occurred (i.e., for constant $\tilde g$ and $\tilde r$).
The value of the exogenous parameters $\tilde a, \tilde g, \tilde r$ are shown above each curve segment.
The regime shifts are visualised by the kinks of the leverage/trust trajectory. They are also visible as jumps in the ROA, ROE plots
shown in panels (b) as a function of leverage. Note that, in each of the panels, 
it can be observed that the return on assets is equal to the return on equity in the steady-state (where $T_{\text{stationary}}=L_{\text{stationary}}$). 

\fig{fig:medium} corresponds to the situation where the intervention (second kink) 
corresponding to an increase of $\tilde g$ and a decrease of $\tilde r$ occurs at an
intermediate time $\tau = \tau_1$. To study the effect of how the timing of intervention affects the steady-state return on assets,
two other cases are studed: \fig{fig:early} corresponds to a central bank intervention occurring earlier (at time $\tau = \tau_0 <\tau_1$) 
while \fig{fig:late} shows the case of an intervention occurring later ($\tau = \tau_2> \tau_1$). 

\begin{figure}[H]
\vspace{-0.5cm}
\centerfloat
\begin{subfigure}{0.6\textwidth}
\includegraphics[trim={0 0.73cm 0 0.6cm},clip,width=\textwidth]{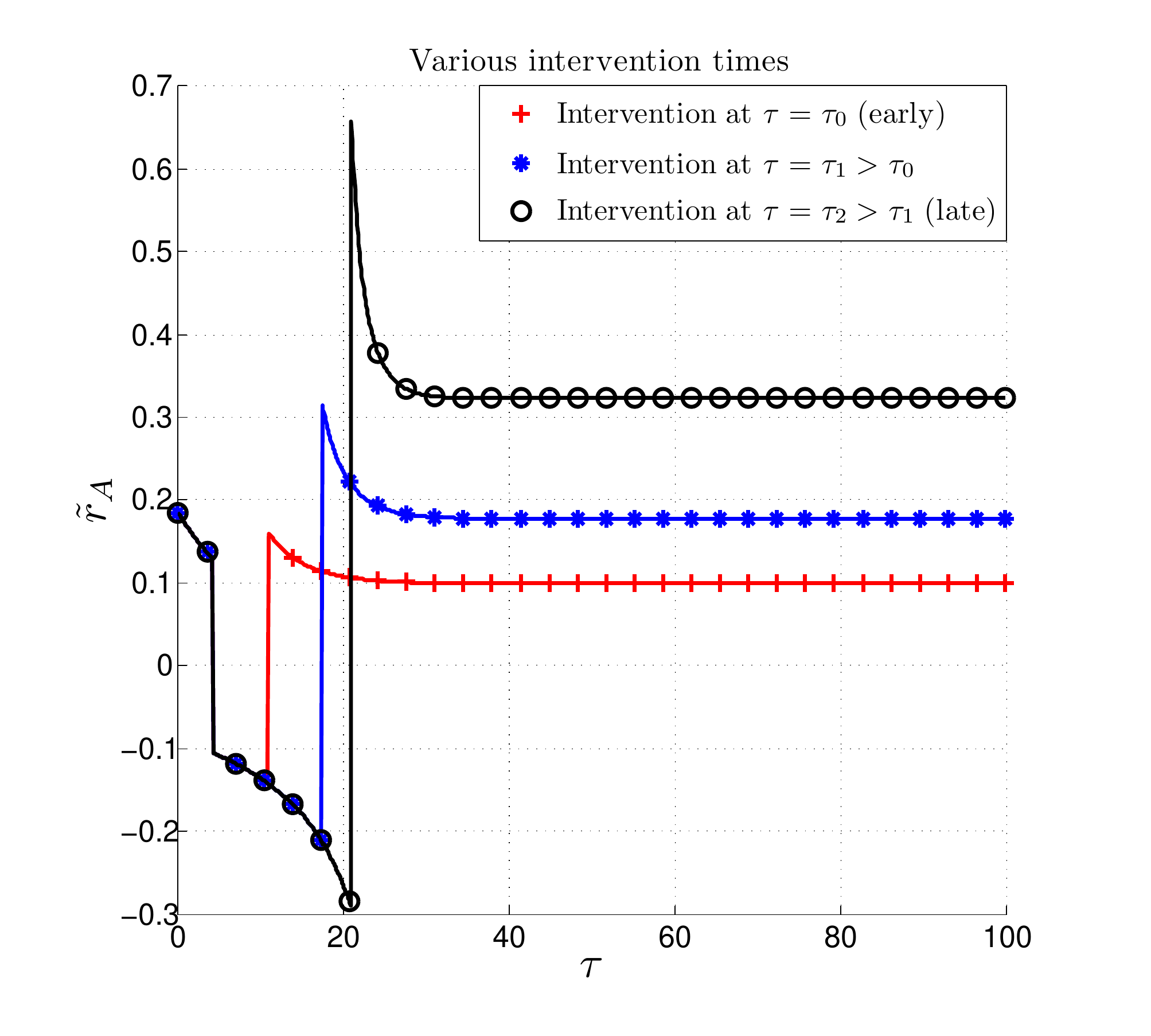}
 \captionsetup{width=0.84\textwidth}
   \vspace{-0.5cm}
\subcaption{The ROA ($\tilde r_A$) as a function of time ($\tau$). Note the overshooting of the ROA, following by a fast
convergence to a stationary level.}
\label{rA_vs_tau}
\end{subfigure}
\begin{subfigure}{0.6\textwidth}
\hspace{0.4cm}\includegraphics[trim={0 0.73cm 0 0.6cm},clip,width=\textwidth]{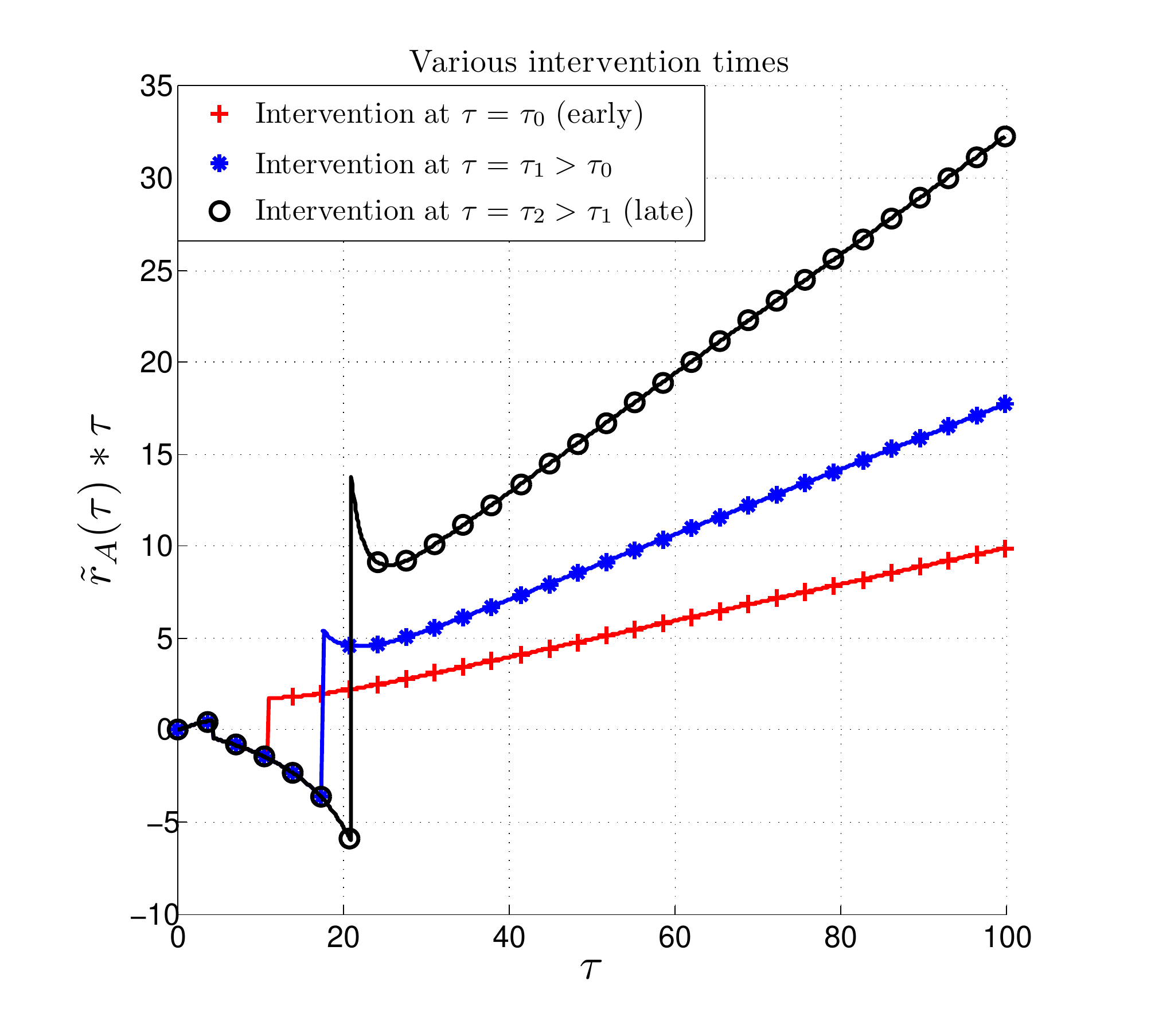}
   \captionsetup{width=0.84\textwidth}
   \vspace{-0.5cm}
\subcaption{$\tilde r_A(\tau)\cdot  \tau$, which represents the natural logarithm of the scaled asset value, as a function of time ($\tau$). }
\label{fig:special_quantity_vs_time}
\end{subfigure}
\caption{Assessment of the efficacy of the central bank's policy intervention and the effect of its timing. The blue curve corresponds to \fig{fig:medium}, the red curve to \fig{fig:early} (earlier intervention) and the black curve to \fig{fig:late} (later intervention).
A late intervention (black curve) results in the highest long-term ROA but comes at the cost of extending the crisis duration.} 
\label{fig:various_times_1plot}
\end{figure}

The effects of the timing of intervention is then summarised in \fig{fig:various_times_1plot}.
Panels (a) and (b) show respectively $\tilde r_A$ and $\tilde r_A(\tau) \cdot \tau$ as a function of $\tau$.
The quantity $\tilde r_A(\tau) \cdot \tau$ can  be interpreted as the natural logarithm of ${A(\tau)}/{A(0)}$ (the scaled asset value at time $\tau$), 
since $A(\tau) = A(0) \e^{\tilde r_A(\tau) \tau}$, which yields $\ln\left( \frac{A(\tau)}{A(0)}\right) = \tilde r_A (\tau) \tau$.
Of course, this holds sensus stricto only for constant $r_A$, so that
$r_A (\tau) \tau$ is only a (good) approximation of  $\ln(A(\tau)/A(0))$.
The main paradoxical conclusion obtained from \fig{fig:various_times_1plot} is that optimising the long-term ROA  comes at the cost of extending the period of crisis.

In the scenarios represented in \fig{fig:medium}-\ref{fig:late}, we have assumed that the first regime change in which $\tilde g$ dropped from a positive value to a negative value is completely exogenous (e.g. due to some external shock). After this first shock, a need for intervention arises. As can be observed, the drop in $\tilde g$ immediately leads to a negative  return on assets and return on equity, and more importantly the trajectory would now head towards a disadvantageous stationary state, i.e. the attractive  fixed point $(T,L)=(1,L_0)$ with a corresponding negative stationary ROA (equal to $-\tilde a$).  In this stationary state, trust exceeds leverage (i.e.  $T_{\text{stationary}} > L_{\text{stationary}}$). This represents an economy in which not all resources are used to their full potential (the economy is ``under-leveraged''). There could be more credit available in the economy to fund profitable projects in the case that the leverage would be  equal to the trust
 (so full potential would be: $T_{\text{stationary}} = L_{\text{stationary}}$). 

Given the fact that intervention in the model economy is needed, the question then arises: what is the optimal time is to intervene. 
Determining the optimal intervention time in our model economy is analogous to studying optimal delays in an
engineering control systems \citep{Fridman14}. Too fast reactions can lead to over-control and unwanted oscillations or instabilities.
Too slow controls may let unwanted regimes to develop.

\fig{fig:various_times_1plot}  shows that the highest steady-state return on assets is obtained in the case of late intervention ($\tau= \tau_2$
shown in figure \ref{fig:late}).
\fig{sf:late_path} elucidates the mechanism driving this result:   postponing the intervention allows for a relatively strong increase in the trust 
variable together with some decrease in leverage, if postponing the intervention long enough (leverage slightly decreases in the last part of the path before intervention in \fig{sf:late_path}). The strong increase in trust is the most notable difference observed when comparing
\fig{sf:medium_path} and \ref{sf:early_path} and it plays a crucial role to attain the highest steady-state return on assets.
This is because it is most effective to intervene when trust is high. Intervening when trust is high ensures 
convergence towards a fixed point $T_{\text{stationary}}=L_{\text{stationary}}$ that is high 
on the diagonal and this translates (for $\tilde g>\tilde r >0$) into a high stationary ROA.

From a dynamical system point of view, that trust increases when $\tilde g$ drops (representing a crisis) results from the fact that
$L_0$ is ``pushed'' inside the domain $L \in [0,1]$ and the trajectories ``bend'' towards the attractive fixed point $(T,L)=(1,L_0)$, which simultaneously means that trust grows. Structurally, 
the increase in trust results from our assumption in the fundamental trust \eq{trust_equation} that trust increases when $T>L$ (and decreases when $T<L$).  In other words, we have engineered an economy in which there is an innate propensity for trust to develop up to its maximum,
allowing in turn leverage to grow so that the economy can attain its full potential of maximum growth.
One could think that there are regimes in which leverage $L$ could grow faster than $T$, overtake it, which would then
lead to a subsequent decrease in trust. However, in the present formulation of our model, this is forbidden
by the ``barrier'' at $T=L$ (the fixed axis) that  cannot be crossed by any trajectory. This property comes from the 
assumption underlying \eq{Debt_dv} in Appendix A that the difference between debt and its maximum available amount
tends to relax exponentially fast with rate $\tilde a$. 
Thus, if at some time, trust is larger than leverage, it can only increase and remain in this region $T \geq L$ at all times.

The above results thus suggest  that the optimal policy intervention strategy is to accept the economic downturn for some time, thereby allowing trust to increase and leverage to decrease somewhat, and then to intervene relatively late by boosting $\tilde g$ and lowering interest rates $\tilde r$.  

One may question whether it makes sense that an economy in crisis can increase trust in the absence of intervention. 
An increasing trust implies that the maximum sustainable leverage increases.
With the lower utilisation of means of production, this might actually reflect a reality. But the psychology of
economic agents in general dominates, with strong risk aversion developing during crisis, leading to 
freezing of capital and under utilisation of resources. Such sub-optimal behaviour can be captured
mathematically by modifying the optimal learning embodied in \eq{Debt_dv} of Appendix A to a weaker one that 
assumes less perfect anticipation of the optimal leverage. Relaxing the rigid structure of \eq{Debt_dv} 
will have the immediate consequence that the axis $T=L$ is no more a fixed axis under all circumstances, 
so that more complex dynamics that can cross it could develop. This will be studied in a subsequent publication.

\section{Calibration of the model and performance} \label{sec:Bayesian}

In this section, the model of the joint dynamics of assets, leverage and trust 
is calibrated to ROE data of the EURO STOXX 50 over the time period 2000-2013, using Bayesian inference (i.e. the Gibbs sampler). 

\subsection{Set-up of the calibration exercise} \label{sec:model_setup}

The specification of the model  for the purpose of calibrating its ROE equation to ROE data is now described.
It consists in an observation equation complemented by state equations, together with the specification of the prior distributions
for the model parameters. The observation equation is the discrete time version of \eq{r_E_expr} to which 
a stochastic residual is added.
The state equations are the discrete time versions of the model equations \eq{L_tau} and \eq{T_tau} for $L$ and $T$ studied previously.
In the state equations, we allow for two regimes to co-exist, corresponding to states $s_1$ and $s_2$ respectively associated
with two distinct values for the parameter $g_s$. The transition between the two regimes is described by a 
standard Markov Switching model with transition matrix $Q$.

 \begin{flushleft}
\begin{align}
& \nonumber \textbf{Observation Equation}\\
&[r_E]_t = g_{s_i} +\frac{L_t}{1-L_t} ( g_{s_i}- r_t) + \epsilon_t,  \hspace{7cm} \epsilon_t \sim \mathcal{N}(0, \sigma_\epsilon^2), \label{r_E_expr_noise}\\ \nonumber 
&\text{where } \;\; t\in \mathbb{Z}^+ \text{ and } s_i=s_i(t)\in \{s_1, s_2\} \text{ indicates the state at time }t.     \\ \nonumber\\ \vspace{-0.3cm}
& \nonumber \textbf{State Equations}\\
& L_{t+1} = L_{t} + (T_{t}- L_{t})    \left(\frac{ g_{s_i}- r_t L_{t}+ a (1-L_{t}) }{1-T_{t}} + k(1-L_{t}) T_{t}  \right) \Delta t ,
 \label{L_Bayesian}\\
&T_{t + 1}= T_{t} +k T_t \left(T_{t}-L_t \right) (1-T_{t}) \Delta t, \label{T_bayesian}\\
& g_{s_i} =  
 \begin{cases} c_1 &\mbox{in } s_1 , \\ 
c_2 & \mbox{in } s_2, 
\end{cases}  \;\;\; \text{(Markov Switching Model).}
  \label{g_AR}  \\
  \nonumber  \\   \vspace{-0.3cm}
 & \nonumber \textbf{Priors} \\
 &\nonumber \sigma_\epsilon^2 \sim \mathcal{IG}(10^{-2},10^{-2}) ,\;\; c_1, \;c_2 \sim \mathcal{U}(-0.25,0.25).\\
 & \nonumber \text{($\mathcal{IG}$ denotes the inverse gamma distribution and $\mathcal{U}$ denotes the uniform distribution).}\\
& \nonumber \text{Transition rate matrix:} \;\;   Q = 
\left[ 
\begin{matrix}
-\lambda & \lambda \\
\mu & - \mu
\end{matrix} \right],  \;\;\;\;\;\;\; \lambda, \mu \sim \mathcal{U}(0,100). \nonumber \\
 &\text{Initial conditions:} \; \; L_1 \sim \mathcal{U}(0.2, 0.3), \; T_1 \sim \mathcal{U}(0.3, 0.4). \nonumber \\ \nonumber \\ \nonumber  
 &\textbf{Parameter values:} \; \;  a = 0.05,\; k=0.05, \; \Delta t = 0.1.
\end{align}
 \end{flushleft}

We sample 200 times and a burn-in sample of  100 is chosen (i.e. the  first 100 samples are discarded). Uninformative priors for the variance of the noise term $\epsilon_t$ and most other parameters ($c_1, c_2, \lambda, \mu$) are chosen. The inverse gamma distribution is commonly used as a prior for the variance  of noise terms.
 Restrictions are imposed on $L_1$ and $T_1$ to ensure that  all variables remain finite and that $T>L$. 
 The parameters that govern time scales ($a, k, \Delta t$) are chosen to be constant and such that all variables remain finite. The time step $\Delta t=0.1$ (year) corresponds to approximately one month, in line with the fact that we supply one data point for $[r_E]_t$ and $r_t$ per month (12 data points per year).
 
We could have chosen to estimate the model parameters $a$ and $k$ as well, as long as $k$ and $a$ remain positive and do not cause variables to become infinite.
 However, estimating $a$ and $k$ (e.g.  $a \sim U(0,10)$ and $k\sim U(0,1)$)  will cause the dynamics of leverage and trust to become somewhat redundant due to the fact that in this way $a$, $k$ will be 
such that leverage and trust very quickly become constant. This would effectively reduce the system of equations that is used for fitting to only \eq{r_E_expr_noise} and \eq{g_AR} with $r_t$ supplied as data and $L_t$ constant.  To preclude that leverage and trust very quickly reach a steady-state,  we assign constant  values to $a$ and $k$. The parameter
$a$ is chosen to be small (and of the same order of magnitude as $g$) to make sure that $g+a$  and thus $L_0$ changes significantly when the state is changed (from $s_1$ to $s_2$ or vice versa). Indeed, recall that $L_0$ controls the position of the attractive fixed point $(1,L_0)$.
This results in  a leverage $L_t$ that fluctuates (decreases in one state, increases in the other state).  Furthermore, $k$ is chosen such that 
$k \cdot \Delta t$ is small, which precludes that $L \to 1$ and $T \to 1$ so as to avoid the divergences associated with the terms
$1/(1-L)$ and $1/(1-T)$ in \eq{r_E_expr_noise} and \eqref{L_Bayesian}.

 The empirical time series $r_t$ is obtained from the Federal Reserve Bank of St. Louis \citep{FRED} and the time series $[r_E]_t $ is calculated based on data obtained from Thomson Reuters of the adjusted closing prices of the EURO STOXX 50. The supplied return on equity values  are monthly averaged yearly returns.

\subsection{ROE and EBITA/Assets ratio: estimated distributions} \label{sec:results_bayes}

\begin{figure}[H]
\centerfloat
\includegraphics[width=\textwidth]{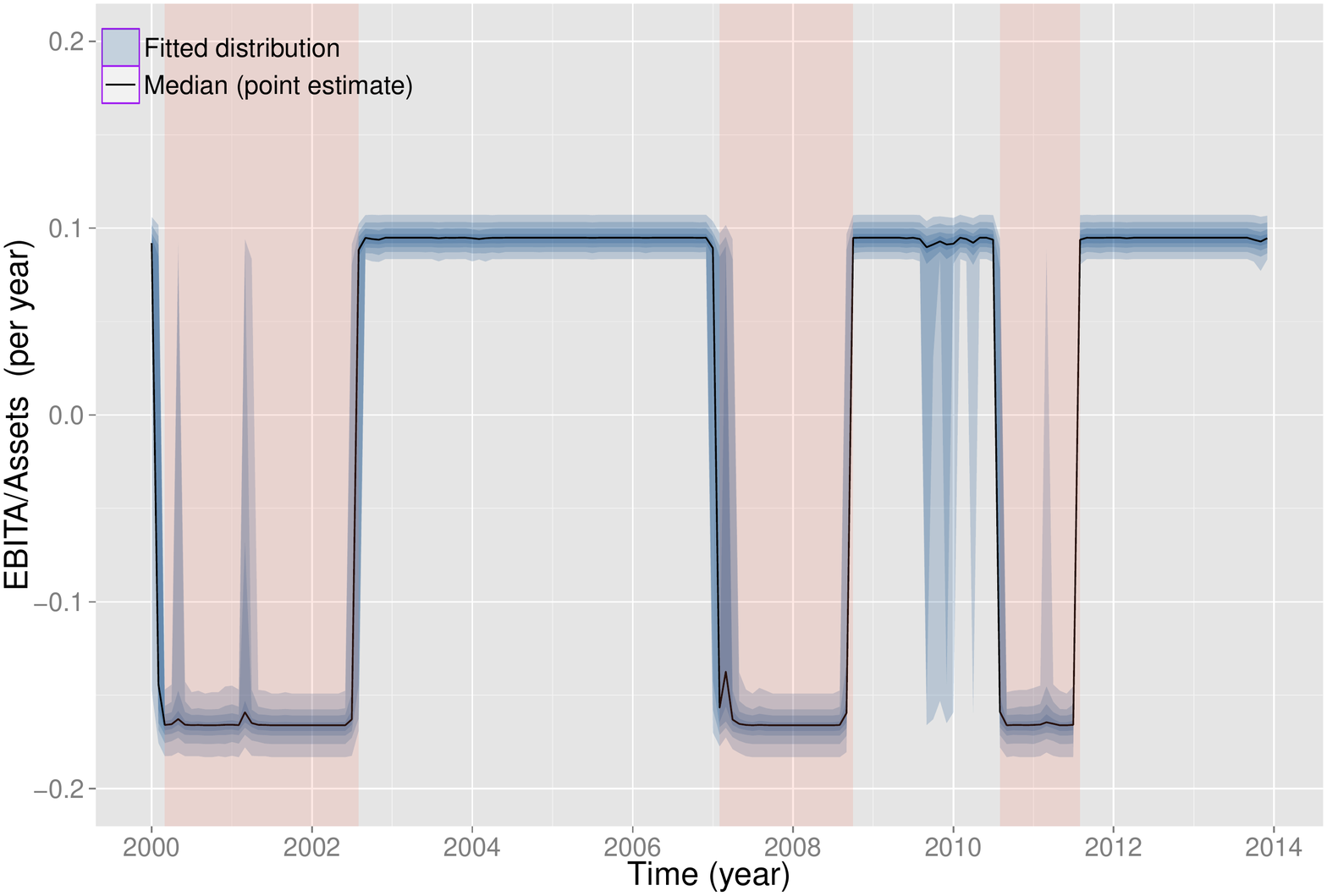}s
\vspace{-.3cm}
\caption{Calibrated distribution of the EBITA/Assets ratio $g$. A Markov Switching Model (MSM) with two states was specified for $g$, which explains the 
observed two levels. The MSM can be viewed as an extended piecewise constant model where the switching between states is endogenized.
 The light red shade is used to distinguish states 1 and 2 and to show the switching events between the states. The probability distribution is shown with different shades of blue: 20\% of the observations  lie within the dark blue region, another 20\% in the lighter blue region that surrounds the dark blue one, and so on.  The black line shows the median (point estimate).  It can be observed that $g$ is negative in the state shaded with light red, while it is positive in the other state.}
\label{fig:g_gARproc}
\end{figure}

\begin{figure}[H]
\centerfloat
\includegraphics[width=\textwidth]{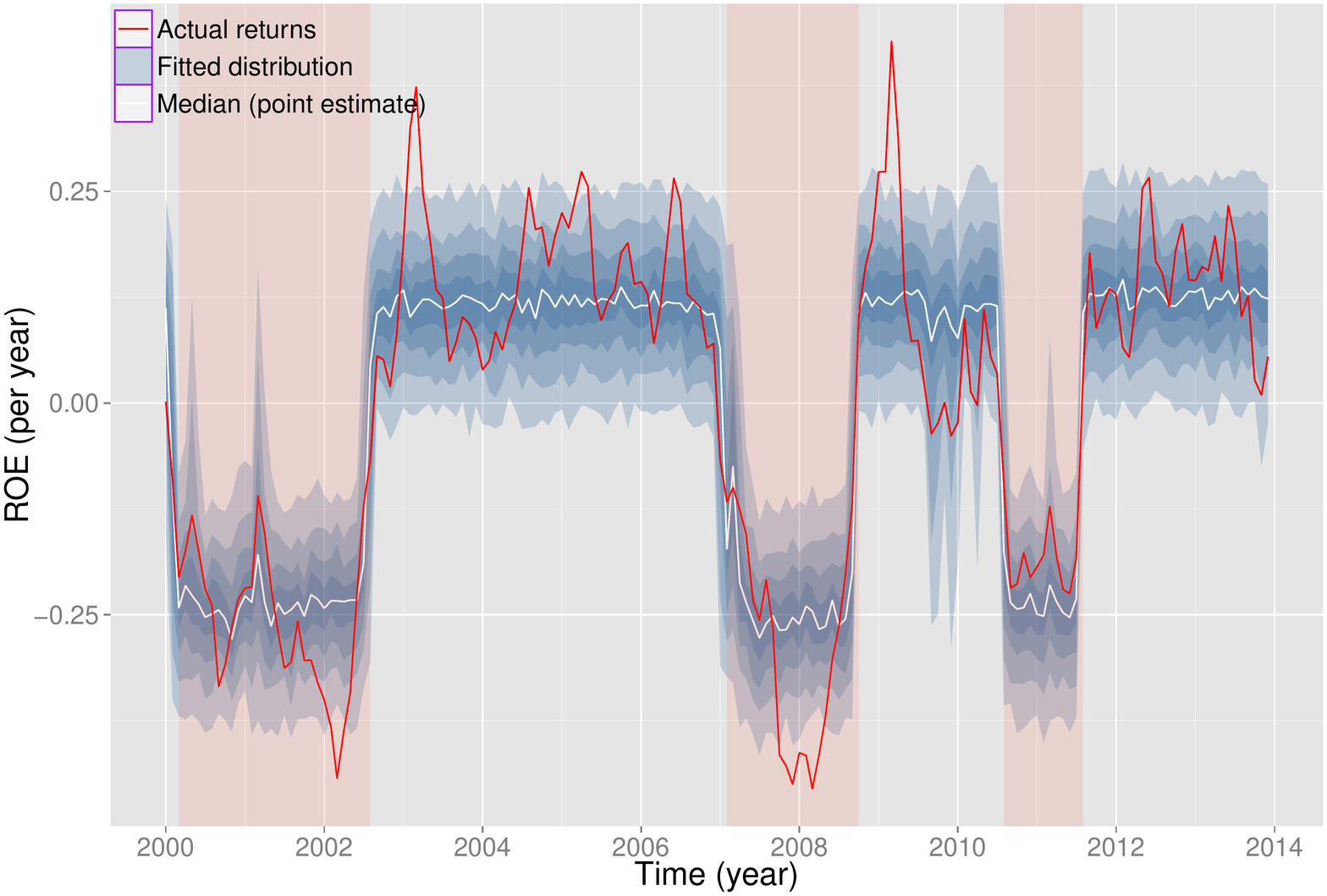}
\vspace{-.3cm}
\caption{Calibrated distribution of the ROE $r_E$. The light red shade is used to distinguish state 1 and 2 and to show the switching events between the states. The states are endogenously determined by the model.  The probability distribution is shown with different shades of blue.  The white line shows the median (point estimate). The red line shows the actual returns of the EURO STOXX 50 (monthly averaged yearly returns) for comparison. It can be observed that the switches between the states occur at the same times as the shocks/jumps in the actual ROE data. The light red shades are seen to correspond  to the periods of negative equity returns (crisis). The first period of crisis corresponds to the dot com crash, the second to the financial crisis of 2007-2008 and the third is associated with the Greek sovereign debt crisis.}
\label{fig:rE_g_AR_proc}
\end{figure}

\fig{fig:g_gARproc} and \ref{fig:rE_g_AR_proc} present the calibrated distribution of the EBITA/Assets ratio $g$ and the return
on equity (ROE) $r_E$ respectively, based on the model specification of \rs{sec:model_setup}. 

 The light red shade is used to distinguish states 1 and 2 and to delineate the switching events between the states.
 The probability distribution is shown with the different shades of blue, that is: 20\% of the observations  lie within the dark blue region, another 20\% in the lighter blue region that surrounds the dark blue one, and so on. The probability distribution of $g$ is narrow and the different shades cannot be (easily) observed.

\fig{fig:rE_g_AR_proc} shows that the model as presented in \rs{sec:model_setup} is successful in fitting the return on equity of the EURO STOXX 50 for the time period 2000-2013. 
It can be observed that the switches between the states, which are   endogenously determined in the model, occur at the same times as the shocks/jumps in the actual ROE data. Furthermore, the light red shade corresponds to the periods of negative equity returns (crisis).  The proposed
dynamics of the EBITA/Assets ratio  $g$ shown in \fig{fig:g_gARproc} is crucial to obtain a good fit of the ROE data and it is clear that the Markov Switching Model with two states is successful.

\subsection{Leverage and Trust: estimated distributions} \label{sec:lever_trust_fittedDistr}

\begin{figure}[H]
\vspace{-0.1cm}
\centerfloat
\begin{subfigure}{0.55\textwidth}
\includegraphics[width=\textwidth]{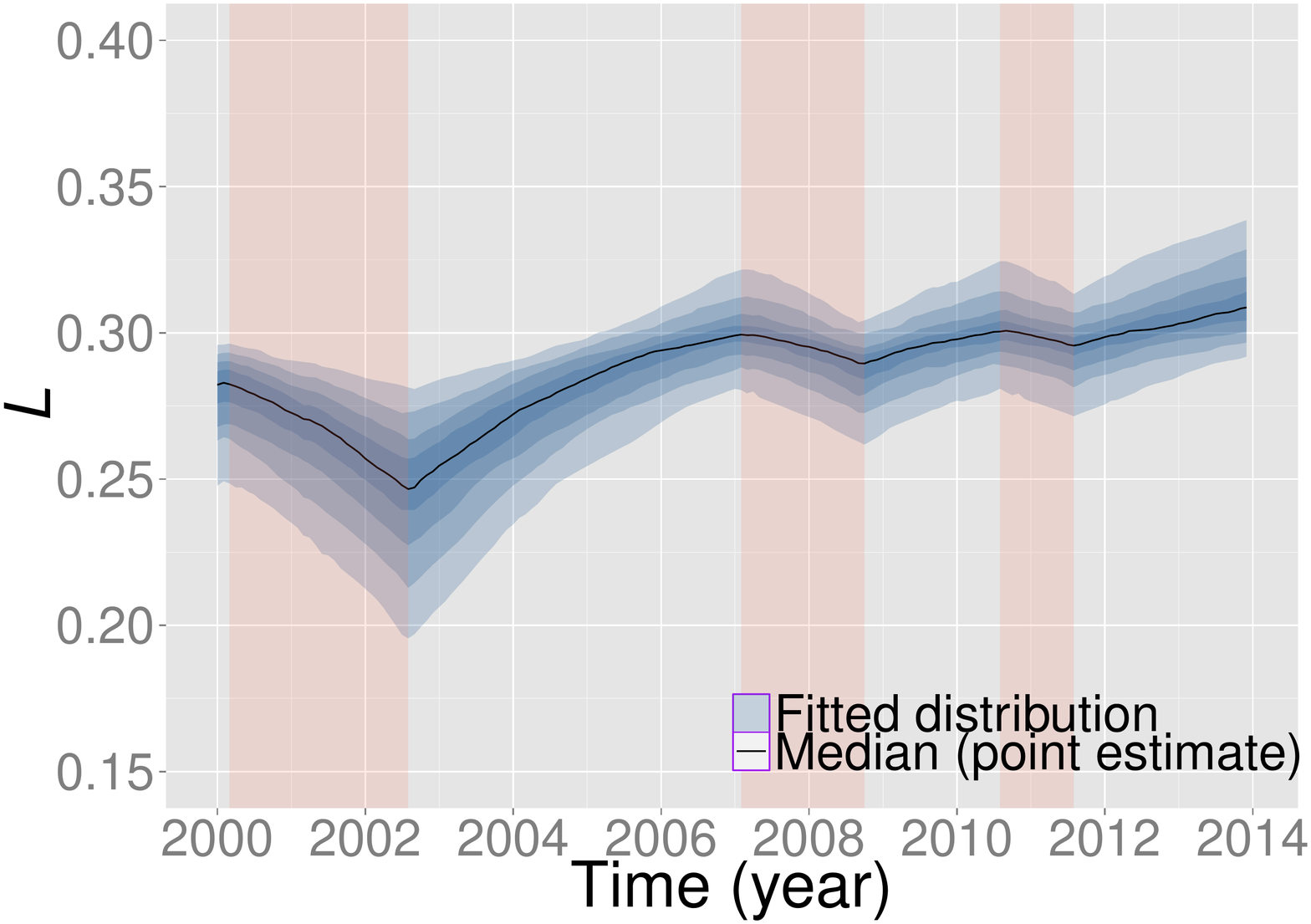} 
\captionsetup{width=0.84\textwidth}
\vspace{-0.5cm}
\subcaption{Calibrated distribution of $L$ as a function of time. The leverage decreases in the state shaded with light red, while it increases in the other state. This results from the fact that $g$ and thus $L_0$ change significantly when the states are switched ($(T,L)=(1,L_0)$ is an attractive fixed point).}
\label{L_g_HMM_new}
\end{subfigure}
\begin{subfigure}{0.55\textwidth}
\includegraphics[width=\textwidth]{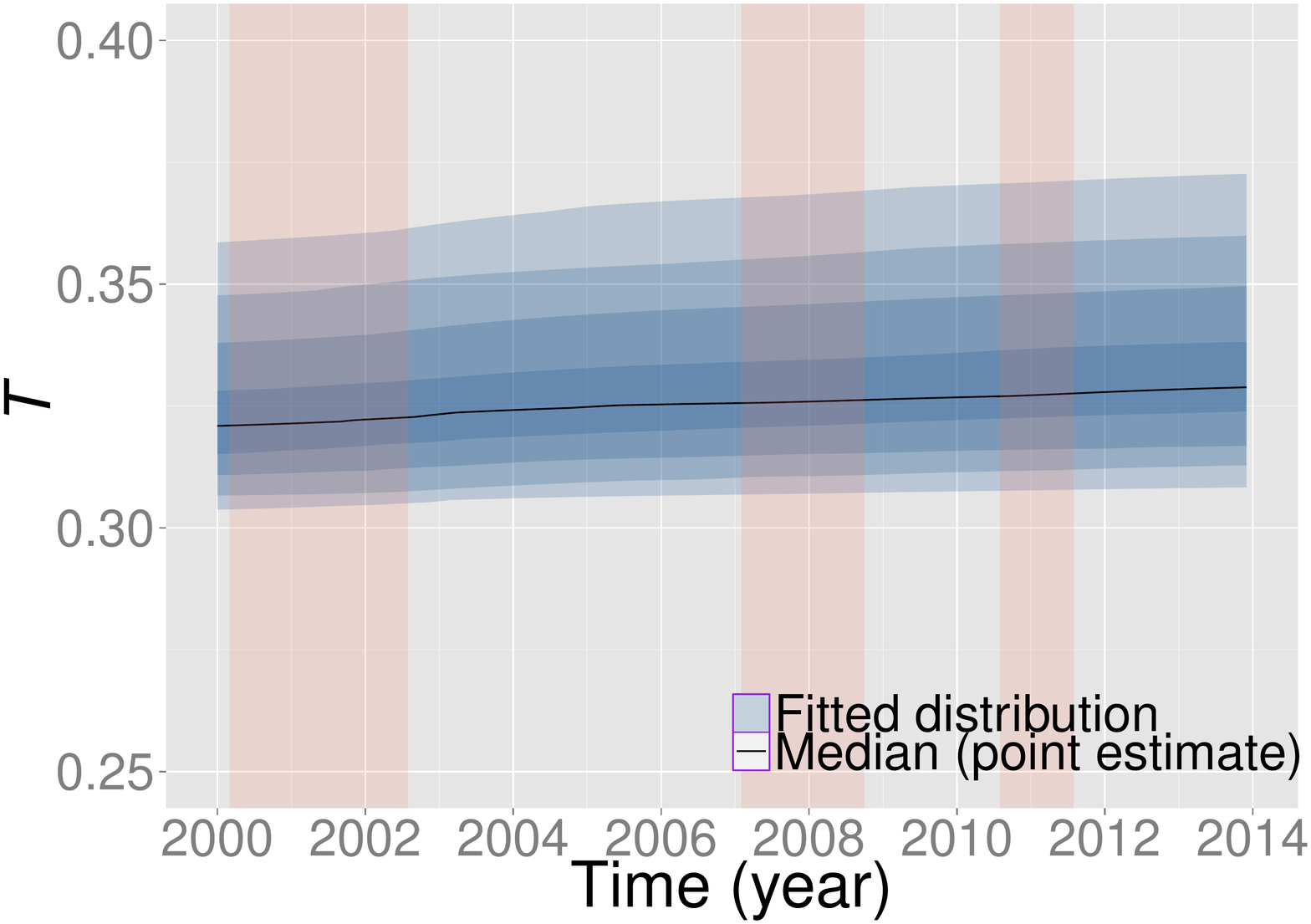} 
   \captionsetup{width=0.84\textwidth}
\vspace{-0.5cm}
\subcaption{Calibrated distribution of $T$ as a function of time. The trust slightly increases to a level determined by 
the values of $k$ and $\Delta t$.}
\label{T_g_HMM_new}
  \end{subfigure}
        \caption{Calibrated distributions of leverage and trust. The light red shade is used to distinguish state 1 and 2 and to show the switching events between the states.   The states are endogenously determined by the model. The probability distributions are shown with the different shades of blue.  The black line shows the median (point estimate). }
  \label{L vs t and T vs L gAR}
  \end{figure}

\fig{L vs t and T vs L gAR} presents the resulting plots of the leverage (panel (a)) and trust (panel (b)) as a function of time,
obtained using the model set-up of \rs{sec:model_setup}. 
\fig{L_g_HMM_new} shows that the leverage decreases in the state shaded with light red (crisis state) and increases in the other state. This is a consequence of the fact that $L_0:=\frac{g+a}{r+a}$ is negative in the state shaded with light red  ($g+a\approx -0.16 + 0.05 = -0.11$ with $g$ taken from \fig{fig:g_gARproc}), while it is positive in the other state ($g+a\approx 0.1 + 0.05 = 0.15$). Recall
that $(T,L)=(1,L_0)$ is an attractive fixed point.

The trust/leverage trajectory lies in the $T>L$ regime as a consequence of the priors chosen for $L_1$ and $T_1$. As discussed previously in \rs{sec:leverage_trust_regime_shifts}, the increase in trust is the  result of the assumption in the fundamental trust equation (\eq{trust_equation}) that trust increases when $T>L$ (and decreases when $T<L$) together with the fact that agents anticipate optimally the evolution of leverage.
This leads to an uncrossable ``barrier''  at the fixed axis $T=L$, so that
trust can only increase when starting from an initial condition with $T>L$.  

\fig{T_g_HMM_new} furthermore shows that trust is almost constant. This is due to the fact that the ``updating term'' in  \eq{T_bayesian} is relatively small ($k \cdot \Delta t = 0.005$), making trust essentially constant.  
\fig{L vs t and T vs L gAR} also illustrates  that $L=1$ and $T=1$ are avoided.
This is important because the model blows up whenever $T\to 1$ or $L \to 1$, since its equations 
exhibit terms proportional to $1/(1-L)$ and $1/(1-T)$ (see \eq{r_E_expr_noise} and \eqref{L_Bayesian}).

\section{Conclusion} \label{sec:conclusion}

A macroeconomic model has been proposed based on the economic variables (i) assets, (ii) leverage and (iii) trust. The main motivation to use these three variables is to focus on the role of credit in the dynamics of economic growth, and to investigate how credit may be associated with both 
economic performance and confidence.
 
Fundamental economic relations and assumptions have been used to describe the joint dynamics of these three variables. The interplay between assets, leverage and trust has been presented in  leverage/trust trajectory plots, accompanied by contour plots of the return on assets.  Several interesting features of the assets, leverage and trust model have been discussed.

The first notable insight is the mechanism of reward/penalty associated with patience, as
quantified by the return on assets.  In regular economies ($g>r$), starting with a trust higher than leverage results in the highest long-term return on assets (which can be seen as a proxy for economic growth). Therefore, patient economies that first build trust and then increase leverage are positively rewarded. We also find that a positive development does not need to be monotonous: before reaching the happy positive
growth steady state: an economy can live through transient regimes during which debt growth exceeds asset growth in the short-run,
before converging to the most favourable long-term state.

Our second main finding concerns a recommendation for the reaction of a central bank to an external shock that 
affect negatively the economic growth. For this,
regime shifts associated with exogenous changes of model parameters have been  studied for 
different leverage/trust trajectories.  The regime shifts represent sudden changes in economic parameters, as a result
of a crisis, or due to the intervention of a central bank. Based on the sample trajectories, 
the effect of the timing of policy intervention has been studied. It was found that late policy intervention in the model economy results in the highest long-term return on assets and largest asset value. Of course, this comes at the cost of suffering longer from the crisis until
the intervention occurs. 
The phenomenon that late intervention is most effective to attain a high long-term return 
on assets can be ascribed to the fact that postponing intervention allows trust to increase first, and it is most effective to intervene when trust is high.

These results derive from our first assumption that trust tends to increase when it is above leverage together with 
our second assumption that economic agents use
an optimal learning embodied in \eq{Debt_dv} of Appendix A of what should be the utilisation of debt
for a given level of trust and amount of assets. Relaxing this with less optimal learning 
may lead to more complex dynamics, which will be studied in a subsequent publication.
 
We have also presented a calibration of the model to empirical data. A calibration set-up has been proposed, 
based on the Euler discretisation of our differential equations governing the dynamics of asset, leverage and trust. 
By specifying a Markov Switching Model for the EBITA/Assets ratio $g$, the model was shown to be very successful in 
fitting the empirical data of the return on equity of the EURO STOXX 50 for the time period 2000-2013.

The fitted distribution of leverage was found to decrease in the state corresponding to crises and to increase in the other 
growing economy state. In the calibrated distribution of the trust variable, it can be observed that there is no time at which trust decreases. This is again a consequence of the assumption in the fundamental trust equation that trust increases when $T>L$ (and decreases when $T<L$)
together with the assumption of optimal learning of the optimal level of debt.

The presented figures also show that the dynamics of leverage and trust can be highly non-monotonous with
curved trajectories, as a result of the nonlinear coupling between the variables. This has an important implication for policy makers, 
suggesting that simple linear forecasting can be deceiving in some regimes and may lead to
inappropriate policy decisions.

\clearpage
\begin{appendices}

\section{Closed-form solution of leverage as a function of trust}  \label{leverage_trust_appendix}

The leverage/trust trajectories  can be solved analytically (closed-form) based on \eq{L_tau} and \eqref{T_tau}. We will outline the derivation and provide the result.

\noindent The first step in deriving the closed form solution is to divide \eq{L_tau} by \eq{T_tau}:
\begin{equation}
\frac{\D L}{\D T} = \frac{1}{T(1-T)}    \left(\frac{\tilde g-\tilde r L+\tilde a (1-L) }{1-T} + (1-L) T \right). \label{dLdT}
\end{equation}

\noindent In terms of $\beta$ and $L_0$, \eq{dLdT} is given by:
\begin{flalign}
\frac{\D L}{\D T}&= \frac{1}{T(1-T)}  \left(\beta\frac{L_0 - L  }{1-T}+ T(1-L)\right), \label{dLdT_beta}\\
 & = \frac{\beta}{T} \cdot \frac{L_0 - 1}{(1-T)^2} + \left[ \frac{\beta}{T (1-T)^2 } + \frac{1}{1-T}\right] \left(1-L\right). \;\; \;\; \; (\beta \neq 0, \pm \infty) \label{dLdT_beta2} 
\end{flalign}

\noindent \eq{dLdT_beta2} is a first order linear differential equation. Solving \eq{dLdT_beta2} yields:
\begin{align}
L(T) &= 1 - K\frac{(1-T)^{1+\beta}}{T^\beta} \e^{-\frac{\beta}{1-T}} 
 +  (L_0 -1)\bigg\{ \left[ \frac{\beta}{1+\beta}+ \frac{1-T}{1+\beta}\right] \notag \\
 & \;\;\;+ \frac{\beta}{1+\beta}  \frac{T^2 }{(1-T)}  \int_0^1{(1-y)^{\beta+1} } {\e^{-\frac{\beta T}{1-T} y } \D y } \bigg\},    \label{LT_final}
\end{align}
where $K$ is an integration constant.

\section{Proof of Theorems  \ref{FixedPoints_determ}-\ref{Theorem_fixedpoints2}}
\label{app_proof_theorem}

\subsection{Determination of fixed points and Jacobian}

\noindent The fixed (stationary) points are the set $(T,L)$ for which $\frac{\D L}{\D \tau} = 0$ and $\frac{\D T}{\D \tau}=0$. From \eq{T_tau} and \eq{L_tau_2} (which are valid for $k\neq 0$), it follows that the axis $T=L$ and the point $(T, L)=(0,L_0)$ satisfy this condition.
Furthermore, $(T,L) =(1,L_0)$ is a fixed point. To show this, consider \eqref{assets1}, \eqref{Debt1}, \eqref{trust_equation} in non-dimensional time for $T=1$:
 \begin{equation}
\begin{dcases}  
& \frac{\D A}{\D \tau}\bigg|_{T=1} = \tilde g A -\tilde r D+ \frac{\D D}{\D \tau},\\ 
 & \frac{\D D}{\D \tau}\bigg|_{T=1} = \tilde a (A- D)+ \frac{\D A}{\D \tau} , \\
     & \frac{\D T}{\D \tau}\bigg|_{T=1}= 0,
\end{dcases}
\label{system_fixed}
\end{equation}
so:  $\frac{\D T}{\D \tau}=0$ (as required to classify as fixed point) and one can substitute the first equation of system \eqref{system_fixed} into the second to find under what condition a solution is admitted:
\begin{flalign}
&\frac{\D A}{\D \tau} = \tilde g A -\tilde r D+ \tilde a (A- D)+ \frac{\D A}{\D \tau}, \\ 
\Leftrightarrow \; & L:= D/A = \frac{\tilde g+\tilde a}{\tilde r+ \tilde a} = \frac{ g+ a}{ r+  a} := L_0. \label{L_admitted}
\end{flalign}

\noindent To summarize, the fixed points are: the axis $T=L$,
the point $(T, L)=(0,L_0)$, and the point $(T, L)= (1,L_0)$.

\noindent Now in order to be able to analyse the stability of the fixed points, define $h :=\frac{\D T}{\D \tau} $ and $g:=\frac{\D L}{\D \tau} $. Furthermore, assume that $(T^*, L^*)$ is an arbitrary fixed point. Now the  linear system for $(T,L)$ close to $(T^*, L^*)$ can be represented by a Taylor expansion around the fixed point:
\begin{flalign}
&\frac{\D T}{\D \tau} = h(T^*, L^*)  + (T-T^*)\cdot \frac{\partial h}{\partial T}\bigg|_{T^*, L^*} + (L-L^*)\cdot \frac{\partial h}{\partial L}\bigg|_{T^*, L^*} + \dots\,, \\
&\frac{\D L}{\D \tau} = g(T^*, L^*)  +(T-T^*) \cdot \frac{\partial g}{\partial T} \bigg|_{T^*, L^*} +(L-L^*) \cdot \frac{\partial g}{\partial L}\bigg|_{T^*, L^*}  + \dots \,.
\end{flalign}
Note that: $\displaystyle h(T^*, L^*)= \frac{\D T}{\D \tau}\bigg|_{T^*, L^*}=0$ and $\displaystyle g(T^*, L^*)= \frac{\D L}{\D \tau}\bigg|_{T^*, L^*}=0$ (the point $(T^*, L^*)$ was defined to be a fixed point), so the resulting system is given by:
{\renewcommand{\arraystretch}{1.5}
\begin{flalign}
{\displaystyle 
\left[
\begin{matrix}
 \frac{\D T}{\D \tau} \\[2ex] \frac{\D L}{\D \tau}
\end{matrix}
\right]
=
{
\left[ 
\begin{matrix}
\frac{\partial h}{\partial T}\big|_{T^*, L^*} & \frac{\partial h}{\partial L}\big|_{T^*, L^*} \\[2ex]
\frac{\partial g}{\partial T}\big|_{T^*, L^*} &\frac{\partial g}{\partial L}\big|_{T^*, L^*}
\end{matrix} \right]}
\left[
\begin{matrix}
T-T^* \\[2ex] L-L^*
\end{matrix}
\right],
}
\end{flalign}
where the 2 by 2 matrix is the Jacobian evaluated at the fixed point $(T^*, L^*)$. \\
}

\noindent  The Jacobian matrix $J$  in its most general variant is given by:
\begin{equation}
J=\left[ 
\begin{matrix}
J_{11} & J_{12} \\
J_{21} & J_{22}
\end{matrix} \right] =
\left[ 
\begin{matrix}
\partial_T \frac{\D T}{\D \tau} & \partial_L \frac{\D T}{\D \tau} \\
\partial_T \frac{\D L}{\D \tau} & \partial_L \frac{\D L}{\D \tau}
\end{matrix} \right],
\label{Jacobian_general}
\end{equation}
which in this case becomes (using \eq{T_tau} and  \eq{L_tau_2}):
\begin{flalign}
J= J(T,L)=
\left[ 
\begin{matrix}
(2T-L)(1-T)-T(T-L) & -T(1-T)\\
(1-L) \left[ \beta \frac{L_0 - L}{(1-T)^2 } + 2 T - L \right] &-\beta \frac{L_0+T-2L}{1-T} - T(1+T-2L)
\end{matrix} \right].
\label{final_jacobian}
\end{flalign}

\noindent The Jacobian in \eq{final_jacobian}  can be used to determine the (linear) stability of the fixed points.

\subsection{$T=L$ analysed: stability and ROA} \label{sec:T_eq_L}

{\bf Stability}

The Jacobian matrix (\eq{final_jacobian}) evaluated at $T=L$ is given by:
\begin{equation}
J(L,L)=
\left[ 
\begin{matrix}
L(1-L) & -L(1-L)\\
(1-L) \left[ \beta \frac{L_0 - L}{(1-L)^2 } +L \right] &-\beta \frac{L_0-L}{1-L} - L(1-L)
\end{matrix} \right].
\end{equation}

\noindent The eigenvalues (denoted $\lambda_1, \lambda_2$) are: $\lambda_1 =0$,\; $\lambda_2 = x+y=-\beta \frac{L_0 - L}{(1-L) }$. \footnote{The eigenvalues of the matrix $J(L,L)$ can be calculated by solving the following equation for $\lambda$:
$\det(J(L,L)- \lambda I) =0,$ where $I$ is the identity matrix. }

\noindent Both eigenvalues are real. If both eigenvalues are negative, then the axis $T=L$ is attractive/stable, while the axis is repulsive/unstable when at least one of the eigenvalues is positive. From $\lambda_2$, it  then follows that (assuming $L\in[0,1]$ and $\beta >0$):
\begin{equation*}
\begin{cases}
\mathrm{For} \; L<L_0, \;  \mathrm{the\; axis}\; T=L\; \mathrm{is \;attractive.}\\
\mathrm{For} \; L>L_0, \; \mathrm{the\; axis}\; T=L \;\mathrm{is\; repulsive.}
\end{cases} 
\end{equation*}
\noindent Note that the above implies that, if $g>r$ (then: $L_0>1$), the axis $T=L$ will be entirely attractive for $L\in [0,1]$.\\

{\bf ROA}

\noindent The corresponding non-dimensional return on assets on the axis $T=L$ is given by:
\begin{flalign}
&\tilde r_{A} |_{T=L} \stackrel{\eqref{r_A_general2}}{=} \tilde g \frac{1}{1-L} -\tilde r \frac{L}{1-L}, \\
\Leftrightarrow \; &\tilde r_{A} |_{T=L} = \tilde g + \frac{L}{1-L} (\tilde g - \tilde r) \stackrel{\eqref{r_E_expr}}{=} \tilde r_E. \label{rE_eq_rA_TL}
\end{flalign}
From \eq{rE_eq_rA_TL}, it follows that on the axis  $T=L$ the return on equity and the return on assets are equal. This result $\tilde r_A = \tilde r_E$ on $T=L$  economically are intuitive. The axis $T=L$ is a steady-state axis and thus describes long-run behaviour. Hence the result can be interpreted as that in the long-run the  return on financial investments (i.e. the return on equity) should equal the growth of the  economy (return on assets). This is in line with the reasoning by \citet{Sornette} and \citet{Dalio}.

By expressing  \eq{rE_eq_rA_TL} as follows:
\begin{flalign}
 &\tilde r_{A} |_{T=L} = \tilde r +(\tilde g - \tilde r) \frac{1}{1-L}, \label{T_equals_L_return}
\end{flalign}
the derivative with respect to $L$ can be easily computed.
The derivative of \eq{T_equals_L_return} with respect to $L$ indicates whether $\tilde r_{A} |_{T=L}$  increases, decreases or stays constant for increasing $L$.  Taking the derivative with respect to $L$ of \eq{T_equals_L_return} gives:
\begin{flalign}
\frac{\D}{\D L}\tilde r_{A} |_{T=L} &= \frac{\tilde g - \tilde r}{(1-L)^2 }, \label{deriv_r_a}
\end{flalign}
which expresses that: if $\tilde g > \tilde r$  ($\tilde g < \tilde r $) then: $\tilde r_{A} |_{T=L}$ increases (decreases) when $L$ increases. When $\tilde g = \tilde r $ then $\tilde r_{A} |_{T=L}$ is  constant  with respect  to $L$. Note that the trajectories can never move on the axis $T=L$ (since points on the axis $T=L$ are in steady-state); however the result is useful to compare the different steady-state points on the axis $T=L$.

\subsection{$(T,L)=(0,L_0)$ analysed: stability and ROA}
{\bf Stability}

The Jacobian matrix (\eq{final_jacobian}) evaluated at $T=0$, $L=L_0$ is given by:
\begin{equation}
J(0,L_0)=
\left[ 
\begin{matrix}
-L_0 & 0\\
-(1-L_0)L_0 &\beta L_0
\end{matrix} \right],
\end{equation}
with eigenvalues  $\lambda_1 = -L_0$, and $\lambda_2 =\beta L_0$.
The eigenvalues are of opposite sign. Hence, the stationary point is a saddle point, which is unstable. 

\hfill 

{\bf ROA}

\noindent The non-dimensional return on assets in the fixed point $(T,L)= (0,L_0)$ is given by:
\begin{flalign}
\tilde r_{A} |_{T=0, L=L_0} &\stackrel{\eqref{r_A_general2}}{=} \tilde g -\tilde r {L_0} - \tilde a L_0  = -\tilde a,
\end{flalign}
which is negative since $\tilde a>0$.

\subsection{$(T,L)=(1,L_0)$ analysed: stability and ROA} \label{sec:attractive_fixed_point_L0}

{\bf Stability}

In order to investigate the stability of the point $(T,L)=(1,L_0)$, it is not possible to consider the Jacobian since this would presume $T \neq 1$ (the Jacobian contains elements $\propto \frac{1}{1-T}$). With perturbation analysis, it is however possible to show that ${(T,L)= (1,L_0)}$ is an attractive fixed point. To do so, consider a small perturbation from the point $T=1$:
\begin{equation}
 T = 1 +\epsilon_{{T}}, \label{Tt}
\end{equation}
where $\epsilon_{\mathrm{T}}$ denotes the perturbation. Then by inserting \eq{Tt} and $L=L_0$ into \eq{T_tau}, the following equation is obtained:
\begin{flalign}
 & \frac{\D \epsilon_{T}}{\D \tau } = -\epsilon_T \cdot \big(\epsilon_T+(1-L_0)\big) (1+\epsilon_T)  \approx -\epsilon_T \cdot \big(\epsilon_T+(1-L_0)\big)  \approx -\epsilon_T \cdot (1-L_0), \label{pert_T}
\end{flalign}
where a first order approximation is taken (the terms proportional to $\epsilon_{T}^2$ or higher order terms are neglected). \\

\noindent By separating variables and integrating \eq{pert_T}, one can find that:
\begin{flalign}
\epsilon_T = C \e^{-(1-L_0) \tau},
\end{flalign}
showing that $\epsilon_T \to 0$ when $\tau \to \infty$ if $L_0<1$, which shows that $T=1$ is attractive if $L_0<1$. \eq{L_admitted} showed that $L=L_0$ is the only solution admitted for $L$, hence:  ${(T,L)= (1,L_0)}$ is an attractive fixed point if $L_0<1$.\\

{\bf ROA}

\noindent To determine the non-dimensional return on assets at the fixed point $(T,L)=(1,L_0)$, note that \eq{r_A_general2} can also be expressed as:
\begin{flalign}
\tilde r_{A} = - \tilde a  + \beta \frac{ L_0 - L}{1-T}+ (T-L) T. \label{r_A_alternative}
\end{flalign}

\noindent In order to be able to take the limit $T \to 1$,  a double integration by parts of \eq{LT_final} can be performed to arrive at:
\begin{flalign}
L(T)&= L_0 - K\frac{(1-T)^{1+\beta}}{T^\beta} \e^{-\frac{\beta}{1-T}} 
   +(L_0 -1)(1-T)  \left\{ -\frac{1}{\beta}  +   \int_0^1 (1-y)^{\beta -1} \e^{-\frac{\beta T}{1-T} y} \D y\right\}. \label{L_FINAL}
\end{flalign}

\noindent One can now substitute \eq{L_FINAL} into \eq{r_A_alternative} to obtain:
\begin{flalign}
\tilde r_{A } = -\tilde a + \beta  \left( K\frac{(1-T)^{\beta}}{T^\beta} e^{-\frac{\beta}{1-T}}
+ (1 - L_0) \left[-\frac{1}{\beta} +  \int_0^1 (1-y)^{\beta-1} e^{-\frac{\beta T}{1-T} y} \, \D y \right]
\right) +  (T-L)T  ,
\end{flalign}
and therefore:
\begin{flalign}
\lim_{T \to 1}  \tilde r_{A} |_{L=L_0} &=- \tilde a -\beta (1-L_0) \frac{1}{\beta}+ (1 - L_0) = -\tilde a. \label{rAT1}
\end{flalign}

\end{appendices}

\clearpage
\def\bibindent{1em}

\end{document}